\documentclass[10pt,twocolumn,twoside]{IEEEtran}
\usepackage{amsmath,amsfonts}
\usepackage{amssymb}
\usepackage{algorithm}
\usepackage{algpseudocode}
\usepackage{siunitx}
\usepackage{enumitem}
\usepackage{booktabs}
\usepackage{multirow} 
\usepackage{color}
\usepackage[noadjust]{cite}
\usepackage{array} 
\usepackage{fontawesome5}
\usepackage{array}
\usepackage{tikz}
\usetikzlibrary{arrows.meta,positioning,fit,calc}

\usepackage[caption=false,font=normalsize,labelfont=sf,textfont=sf]{subfig}
\usepackage{textcomp}
\usepackage{stfloats}
\usepackage{url}
\usepackage{verbatim}
\usepackage{graphicx}
\def\BibTeX{{\rm B\kern-.05em{\sc i\kern-.025em b}\kern-.08em
    T\kern-.1667em\lower.7ex\hbox{E}\kern-.125emX}}
\usepackage{balance}

\makeatletter
\renewcommand*{\ALG@name}{Protocol}
\makeatother

\newcommand{\bbR}{\ensuremath{{\mathbb R}}}
\newcommand{\bbZ}{\ensuremath{{\mathbb Z}}}
\newcommand{\bbN}{\ensuremath{{\mathbb N}}}

\newcommand{\calC}{{\mathcal C}}
\newcommand{\calD}{\mathcal{D}}

\newcommand{\calX}{\mathcal{X}}

\newcommand{\calN}{\mathcal{N}}
\newcommand{\calV}{\mathcal{V}}
\newcommand{\calE}{\mathcal{E}}
\newcommand{\calG}{\mathcal{G}}
\newcommand{\calO}{\mathcal{O}}

\newcommand{\calK}{\mathcal{K}}
\newcommand{\calF}{\mathcal{F}}
\newcommand{\calZ}{\mathcal{Z}}

\newcommand{\sfx}{\mathsf{x}}
\newcommand{\sfy}{\mathsf{y}}

\newcommand{\sfL}{\mathsf{L}}
\newcommand{\sfQ}{\mathsf{Q}}

\newcommand{\Share}{\mathsf{Share}}

\newcommand{\ini}{\mathsf{ini}}
\newcommand{\avg}{\mathsf{avg}}

\newcommand{\msg}{\mathsf{Msg}}

\newcommand{\Sim}{\mathsf{Sim}}
\newcommand{\RMSE}{\mathsf{RMSE}}

\newcommand{\bfk}{\mathbf{k}}
\newcommand{\bfK}{\mathbf{K}}

\newcommand{\bfy}{\mathbf{y}}

\newcommand{\bfzero}{\mathbf{0}}
\newcommand{\bfone}{\mathbf{1}}

\newtheorem{example}{Example}
\newtheorem{asm}{Assumption}
\newtheorem{lem}{Lemma}
\newtheorem{prop}{Proposition}
\newtheorem{thm}{Theorem}
\newtheorem{cor}{Corollary}
\newtheorem{defn}{Definition}
\newtheorem{rem}{Remark}

\newcommand{\modp}{~\mathrm{mod}~}

\begin{document}
\title{Privacy-Preserving Fully Distributed Gaussian Process Regression}
% \author{Anonymous Authors}

\author{Yeongjun Jang, Kaoru Teranishi, Jihoon Suh, and Takashi Tanaka% <-this % stops a space
\thanks{*This work was supported by the National Research Foundation of Korea (NRF) grant funded by the Korea government (MSIT) (No. RS-2024-00353032 and RS-2026-25504174) and by the JSPS KAKENHI Grant Number JP26K00966. (\textit{Corresponding author: Yeongjun Jang})
}
\thanks{Y.~Jang is with ASRI, Department of Electrical and Computer Engineering, Seoul National University, Seoul, 08826, Korea (email: jangyj0512@snu.ac.kr).
}
\thanks{K.~Teranishi is with the Department of Information and Physical Sciences, Graduate School of Information Science and Technology,
The University of Osaka, Osaka, 565-0871, Japan (email: k-teranishi@ist.osaka-u.ac.jp).
}
\thanks{J.~Suh and T.~Tanaka are with the School of Aeronautics and Astronautics, Purdue University, West Lafayette, IN 47907, USA (email: \{suh95,tanaka16\}@purdue.edu).
}
\thanks{T.~Tanaka is also with the Elmore Family School of Electrical and Computer Engineering, Purdue University, West Lafayette, IN 47907, USA.
}
% <-this % stops a space
}

\markboth{Journal of \LaTeX\ Class Files,~Vol.~18, No.~9, September~2020}%
{How to Use the IEEEtran \LaTeX \ Templates}

\maketitle

\begin{abstract}
Although distributed Gaussian process regression (GPR) enables multiple agents to jointly learn a model of the target function, its collaborative nature poses a risk of private data leakage. 
To address this, we propose a privacy-preserving fully distributed GPR protocol based on secure multi-party computation, which hides each agent's individual contribution from semi-honest coalitions of bounded size, beyond what is implied by the aggregated value.
Building upon a secure distributed average consensus algorithm, it guarantees that each agent's local model practically converges to the same global model obtained by the standard distributed GPR.  
Formal privacy guarantees are established within the simulation based security paradigm.
The protocol is further extended to privacy-preserving optimization of kernel hyperparameters, which is critical yet often overlooked in the literature.
Experimental results demonstrate the effectiveness and practical applicability of the proposed method.
\end{abstract}


% \begin{itemize}
%     \item \cite{LeedKimj20,ZhaoSunh22,LiruStur18,BonaIvan17,YuanSunh25,Sunh26,ChouKlie26}
%     \item \cite{LeedKimj20}: Additively HE. Leader agent (server) 
%     \item \cite{ZhaoSunh22,LiruStur18,BonaIvan17,YuanSunh25,Sunh26,ChouKlie26}: Secure aggregation in the context of federated learning. Require a central server 
% \end{itemize}

\section{Introduction}

Gaussian process regression (GPR) is a Bayesian nonparametric regression method that effectively models complex functions from observed data \cite{WillRasm06}.
An appealing feature of GPR is its ability to provide practical and reliable regression error bounds \cite{FiedSche21}, which makes it particularly attractive for control applications that require uncertainty quantification.
Consequently, GPR has been utilized in areas such as safe control \cite{HewiKabz19,JangChan24,ChanLane26} and state estimation \cite{LeekJohn17,LiqiHong24}.

In many real-world scenarios, data are naturally distributed across multiple agents, for example, across mobile devices in which user information is stored locally.
This has motivated the development of distributed GPR algorithms \cite{RasmGhah01,DeisNgju15}, where each agent independently performs regression on its local dataset, and the resulting local models are combined by a central aggregator or through a distributed protocol.
Such approaches can improve predictive accuracy by incorporating heterogeneous information in the local datasets, and achieve scalability by distributing computational burden across the participating agents.

Despite these advantages of distributed GPR, its collaborative nature introduces new privacy concerns.
For example, each agent's local dataset may contain sensitive information that must remain confidential, either due to privacy regulations (e.g., patient health records or financial transaction data) or because the dataset is proprietary (e.g., industrial process data).
Even when raw data are not directly exchanged, the exchanged local models might inadvertently disclose some information about the underlying dataset.

In the context of distributed GPR, there have been attempts to address this issue by employing cryptographic techniques such as secure multi-party computation (SMPC) and homomorphic encryption (HE), as well as the notion of differential privacy (DP) (see Section~\ref{subsec:related} for a review of related work).
However, to the best of our knowledge, existing SMPC and HE based methods require a trusted central server to perform aggregation, which may be unrealistic when the participating agents mutually distrust each other.
Moreover, such an architecture imposes substantial computation and communication burdens on the central server and creates a single point of failure.
On the other hand, DP based methods ensure privacy by deliberately injecting noise, which inevitably leads to a trade-off between privacy and accuracy.

To overcome these limitations, we propose a \textit{privacy-preserving fully distributed GPR protocol} based on SMPC that hides each agent's individual contribution from semi-honest\footnote{A semi-honest agent follows the protocol as specified but may attempt to learn additional information from the messages it receives.
} coalitions of bounded size, beyond what is already implied by the aggregated value.
Specifically, we make the following key contributions:
\begin{itemize}[leftmargin=*]
    \item Building on a secure distributed average consensus algorithm developed herein, the protocol enables each agent's local model to practically converge to the global model that would be obtained by the standard distributed GPR.
    The consensus algorithm adopts the secure aggregation scheme of \cite{AlexDaru19}, while improving upon it in that neither encryption of the coupling weights nor a trusted central server is required.
    \item Since SMPC operates over integers, the protocol requires quantization of real-valued quantities. 
    We analyze the effect of quantization errors on convergence and derive a sufficient condition on the parameters under which practical convergence is guaranteed.
    By refining the parameters, the convergence error can be made arbitrarily small without compromising privacy, at the cost of increased communication cost.
    \item  We establish formal privacy guarantees using the simulation based security paradigm of \cite{Gold01}, in which the agents are modeled as semi-honest but potentially colluding.
    \item We extend the proposed protocol to privacy-preserving optimization of kernel hyperparameters, which is essential for prediction accuracy but has not been explored within a privacy-preserving framework.
    \item Experimental results on two real-world benchmark datasets show that the proposed protocol attains a substantially lower runtime than the methods of \cite{LuojZhan23} and \cite{NawaChen24}, demonstrating its effectiveness and practical applicability.
\end{itemize}

\subsection{Related work}\label{subsec:related}

There has been a growing interest in developing privacy-preserving GPR algorithms, and existing works can largely be classified into methods based on SMPC, HE, and DP.

In \cite{LuojZhan23}, multiple trusted central servers run an SMPC protocol to aggregate local datasets and train a single centralized GPR model.
This approach can suffer from scalability issues, as the servers should handle computationally costly operations, such as matrix inversion and exponentiation.
In contrast, our approach alleviates this burden by aggregating locally computed models instead.

On the other hand, \cite{NawaChen24} and \cite{FennPyze20} proposed methods based on HE, a cryptographic primitive that enables direct computation on encrypted data without decryption.
In \cite{FennPyze20}, a server holds a privately pre-trained GPR model, while a client submits an encrypted test input and receives the corresponding prediction without revealing the input to the server. 
This enables privacy-preserving prediction for a fixed model but does not support collaborative model training.
To address this limitation and improve scalability, \cite{NawaChen24} implemented a distributed GPR algorithm using multi-key HE. 
Nonetheless, their framework relies on a trusted central server for aggregation and decryption, and the computational complexity of multi-key HE can limit its practical applicability.

Several works have applied DP to GPR by injecting carefully designed noise into intermediate computations or prediction outputs to guarantee privacy \cite{HallRina13,SmitAlva18,HonkMelk21}, but this inevitably leads to a trade-off between accuracy and privacy.

Importantly, none of the aforementioned works have incorporated hyperparameter optimization into their privacy-preserving frameworks, even though it is essential for achieving accurate predictions in GPR \cite{WillRasm06}.

% More broadly, the problem of securely aggregating local values without revealing individual contributions has been extensively studied under the name of secure aggregation \cite{LeedKimj20,LiruStur18,ZhaoSunh22,BonaIvan17,ChouKlie26,SavsPyrg21,MayiWood23}.
% Although our problem is related to it, our contribution is distinct in that we focus on the fully distributed training of GPR models and provide theoretical analyses tailored to this problem.
% Moreover, most existing works on secure aggregation have been developed in the context of federated learning \cite{ZhaoSunh22,BonaIvan17,ChouKlie26,SavsPyrg21,MayiWood23} and rely on a trusted central server for aggregation, making their direct application to our problem nontrivial.

\subsection{Notations}
The sets of integers, positive integers, non-negative real numbers, positive real numbers, and real numbers are denoted by $\bbZ$, $\bbN$, $\bbR_{\ge 0}$, $\bbR_{>0}$, and $\bbR$, respectively.
The identity and the zero matrix are denoted by $I$ and $\bfzero$, respectively, with their dimensions indicated as subscripts when necessary.
For a (matrix) vector of scalars, $\|\cdot\|$ denotes the (induced) infinity norm.
The cardinality of a finite set $A$ is denoted by $|A|$.
For a sequence $v_1,\ldots,v_n$ of scalars or vectors, we define $[v_1;\cdots;v_n]:=[v_1^\top, \cdots, v_n^\top]^\top$.

\section{Preliminaries and Problem Formulation}\label{sec:problem}

\subsection{Distributed Gaussian process regression}\label{subsec:DGPR}

We briefly introduce the distributed Gaussian process regression (GPR) algorithm of \cite{DeisNgju15}.
Consider a network of $M\in\bbN$ agents,
where each agent $i \in \{1,\ldots,M\}$ is equipped with a local dataset $\calD_{i}$ consisting of noisy observations of an unknown function $f:\bbR^n\to \bbR$:
\begin{equation}\label{eq:dataset}
    \calD_{i} \!:=\! \{(x_{i_l},y_{i_l}) \mid y_{i_l} = f(x_{i_l})+\epsilon_{i_l}, ~ l=1,\ldots,N_i\}.\!\!
\end{equation}
Here, $N_i\in\bbN$ is the size of $\calD_i$, and measurement noise $\epsilon_{i_l}\in\bbR$ follow an i.i.d. zero-mean Gaussian distribution with variance $\sigma_\epsilon^2>0$.
For simplicity, we focus on scalar-valued functions, as vector-valued functions can be handled by learning each output dimension separately.

We assume that the agents share a prior Gaussian process (GP) specified by a prior mean function $m:\bbR^n\to \bbR$ and a prior covariance function (or kernel) $k:\bbR^n\times\bbR^n \to \bbR$.
Without loss of generality, we consider a zero prior mean function $m\equiv 0$ in this paper, and assume that the kernel $k$ is positive definite.\footnote{The kernel $k$ is said to be positive definite if the matrix $\bfK_i$ in \eqref{eq:yi} is positive semidefinite for any $N_i\in\bbN$ and $x_{i_l}\in\bbR^n$ for $l=1,\ldots,N_i$.} 
By conditioning the prior GP on its local dataset, each agent $i$ obtains a local posterior mean function $\hat{f}_{i}:\bbR^n\to \bbR$ and posterior variance function $V_{i}:\bbR^n\to \bbR$. 
These functions serve as the agent's estimate of the unknown function $f$ and an uncertainty measure of the estimate, respectively. 
For a given test point $x\in\bbR^n$, they are explicitly computed as
\begin{equation}\label{eq:posterior}
    \begin{split}
    \!\!\hat{f}_i(x) &:= \bfk_i^\top(x)(\bfK_i+\sigma_\epsilon^2\cdot I_{N_i})^{-1}\bfy_i, \\
    \!\!V_i(x) &:= k(x,x)-\bfk_i^\top(x)(\bfK_i+\sigma_\epsilon^2\cdot I_{N_i})^{-1}\bfk_i(x), 
    \end{split}
\end{equation}
where 
\begin{align}\label{eq:yi}
    \bfk_i(x)&:=
    \begin{bmatrix}
            k(x,x_{i_1}) \\
            \vdots \\
            k(x,x_{i_{N_i}})
    \end{bmatrix}\in\bbR^{N_i}, ~~ \bfy_i:=
    \begin{bmatrix}
        y_{i_1} \\ \vdots \\ y_{i_{N_i}}    
    \end{bmatrix}\in\bbR^{N_i}, \\
    \bfK_i&:=
    \begin{bmatrix}
        k(x_{i_1},x_{i_1}) &\!\! \cdots &\!\! k(x_{i_1},x_{i_{N_i}}) \\
        \vdots &\!\! \ddots &\!\! \vdots \\
        k(x_{i_{N_i}},x_{i_1}) &\!\! \cdots &\!\! k(x_{i_{N_i}},x_{i_{N_i}})
    \end{bmatrix} \!\!\in \bbR^{N_i \times N_i}. \nonumber
\end{align}

The product-of-experts variant of the distributed GPR algorithm in \cite{DeisNgju15} aggregates \eqref{eq:posterior}, and computes a collaborative estimate $\hat{f}(x)$ and an uncertainty measure $V(x)$ as follows: 
\begin{equation}\label{eq:DGPR}
    \begin{split}
        \hat{f}(x) &:= V(x)\sum_{i=1}^M V_i^{-1}(x) \cdot \hat{f}_i(x), \\
    V(x) &:= \frac{1}{\sum_{i=1}^M V_i^{-1}(x)}.
    \end{split}
\end{equation}
It can be seen from \eqref{eq:DGPR} that an estimate $\hat{f}_i(x)$ with smaller $V_i(x)$, i.e., when agent $i$ is confident, is assigned a greater weight.
Moreover, positive definiteness of the kernel $k$ guarantees $V_i(x) > 0$ for all $x$ \cite[Section~2]{WillRasm06}, ensuring that \eqref{eq:DGPR} is well-defined. 

Distributed GPR alleviates the poor scalability of standard centralized GPR. 
The biggest computational overhead lies in computing the matrix inversions associated with \eqref{eq:posterior}, whose computational complexity is given by $\calO(N_i^3)$.
In the centralized implementation, all agents' datasets are collected and a single square matrix of dimension $\sum_{i=1}^M N_i$ is formed analogously to $\bfK_i$, whose inversion incurs a computational cost of $\calO((\sum_{i=1}^M N_i)^3)$. 
In contrast, the distributed counterpart in \eqref{eq:DGPR} incurs only $\calO(\sum_{i=1}^M N_i^3)$.

\begin{rem}\upshape
For the simplicity of presentation, we focus on the product-of-experts variant given by \eqref{eq:DGPR}, which tends to be accurate when the local datasets are complementary and at least one agent is informative at the test point $x$, i.e., $V_i(x) \ll k(x,x)$ for some $i$ \cite[Section~3.2.1]{DeisNgju15}.
It is emphasized, however, that the proposed framework can be extended in a straightforward manner to more sophisticated variants, such as the generalized product-of-experts and (robust) Bayesian committee machine introduced in \cite{DeisNgju15}.
\end{rem}

\subsection{Problem formulation}\label{subsec:problem}

Although distributed GPR offers computational advantages over standard centralized GPR, its implementation introduces new privacy concerns. 
For example, each agent's local dataset may contain sensitive data that must be kept private.
However, because the aggregation step in \eqref{eq:DGPR} involves local posterior means and variances that are computed from these local datasets, it may lead to indirect leakage of such sensitive data.
We provide a practical scenario in which such privacy concerns become relevant, through the following example.

\begin{example}\upshape\label{ex}
    Consider a group of hospitals that are connected over a network topology. 
    One of the hospitals receives a new patient and wishes to estimate the health risk or the effectiveness of a certain treatment based on the data of previous patients.
    To improve the quality of the estimate, it queries other hospitals to collaborate using the distributed GPR algorithm.
    However, due to privacy regulations, hospitals should not disclose information about their patients. 
    In this setting, directly sharing the local posterior mean and variance may not be acceptable, as it may reveal some information about the patients.
    Moreover, assuming the presence of a trusted central server could be unrealistic, as hospitals may belong to competing institutions or operate in different countries.
\end{example}

% \begin{example}
% % [Joint Portfolio Optimization]
% \upshape\label{ex2}
% The principle of Model Predictive Control (MPC)---where a planning problem is solved over a finite horizon but only the current action is executed---has been widely used in financial applications such as dynamic option hedging \cite{bemporad2014dynamic} and multi-period trading strategies \cite{boyd2017multi}. 
% Consider a scenario in which multiple firms collaborate using the distributed GPR algorithm to construct a model for such tasks. 
% Each firm may have specialized expertise in a particular asset class or uneven access to alternative datasets,
% giving it unique advantages in forecasting future risks and expected returns. 
% However, in order to maintain their competitive advantages, the firms wish to keep their local datasets $\calD_i$ private. \hfill $\square$

% \end{example}

Motivated by such privacy concerns, we aim to design a privacy-preserving fully distributed GPR protocol based on secure multi-party computation (SMPC). 
The protocol should enable each agent to obtain \eqref{eq:DGPR}, while hiding each agent's individual contribution $\hat{f}_i(x)$ and $V_i(x)$ from semi-honest coalitions of bounded size, beyond what is already implied by \eqref{eq:DGPR} itself.

The problem setting is formalized below.
Let us consider a multi-agent system, where each agent is equipped with a dataset of the form \eqref{eq:dataset}. 
The network topology of the system is modeled by a graph $\calG=(\calV,\calE)$, where $\calV:=\{1,2,\ldots, M\}$ is a finite nonempty set of $M\in\bbN$ agents, and $\calE \subseteq \calV \times \calV$ is an edge set of ordered pairs of agents. 
We exclude self-connection, i.e., $(i,i)\notin \calE$.
An edge $(i,j)\in\calE$ indicates that agent $i$ can directly send information to agent $j$.
The set of neighbors of agent $i\in\calV$ is defined as $\calN_i:=\{j\in\calV \mid (j,i)\in\calE\}$.
We assume that the graph $\calG$ is undirected and connected---that is, $(i,j)\in\calE$ if and only if $(j,i)\in\calE$, and for every pair of agents $i,j\in\calV$ there exists a finite sequence of agents $(v_0=i,v_1,\ldots,v_l=j)$ such that $(v_r,v_{r+1})\in\calE$ for $r=0,1,\ldots,l-1$.

% Before proceeding, we impose the following structural assumption on the graph $\calG$. 

% % \begin{asm}\upshape\label{asm:graph}
% %     Each agent $i\in\calV$ knows the agents that are within two hops away from itself. 
% %     For every edge $(i,j)\in\calE$, agents $i$ and $j$ share at least one common neighbor.
% % \end{asm}

% The first condition in Assumption~\ref{asm:graph} is mild, as it only requires local knowledge of the graph. 
% The second condition strengthens the graph's connectivity, which is often necessary for ensuring resilience and fault tolerance of distributed protocols \cite{SundHadj10,PasqBicc11}.

\section{Secure Distributed Average Consensus}\label{sec:avg}

We first present a secure fully distributed average consensus protocol based on additive secret sharing, which is a widely used cryptographic technique for secure multi-party computation (SMPC) \cite{CramDamg15}. 
Then, we formally establish the protocol's security using the simulation based security paradigm of \cite{Gold01}.
This protocol will serve as the main building block of our privacy-preserving fully distributed Gaussian process regression (GPR) protocol.

\subsection{Additive secret sharing}\label{subsec:ss}

Additive secret sharing splits a given message into random-like shares that individually reveal no information about the message.
It is defined over the finite set of integers $\bbZ_q:=\bbZ\cap[-q/2,q/2)$, where $q\in\bbN$ is the modulus.
The modulo operation defined by $a\modp q:= a- \lfloor (a+q/2)/q \rfloor q\in\bbZ_q$ for all $a\in\bbZ$ maps $\bbZ$ onto $\bbZ_q$.
For vectors, it is applied componentwisely.

The share generation algorithm 
\begin{equation*}
     (s_1,\ldots,s_n) \leftarrow \Share(m,n) 
\end{equation*}
splits a message $m\in\bbZ_q^p$ into $n\in\bbN$ shares $(s_1,\ldots,s_n)$, where $s_1,\ldots,s_{n-1}\in\bbZ_q^p$ are sampled uniformly at random from $\bbZ_q^p$ and $s_n:=m-\sum_{l=1}^{n-1}s_l\modp q\in\bbZ_q^p$.
By construction, the message $m$ can be recovered as $ m = \sum_{l=1}^n s_l\modp q$.
However, if $n-1$ (or fewer) shares among $(s_1,\ldots,s_n)$ are given, no information about the message $m$ can be inferred due to the randomness of the shares.

\subsection{Average consensus with quantized communication}\label{subsec:metro}

Since additive secret sharing is defined over integers rather than real numbers, we consider the average consensus problem in which transmitted signals are quantized to integers.
% This will serve as the basis for the proposed secure distributed average consensus protocol.

Accordingly, let the state $z_i(t)\in\bbR^p$ of each agent $i\in\calV$ be quantized as
\begin{equation}\label{eq:Lz}
        \sfQ(z_i(t)) := \left\lceil \frac{z_i(t)}{\sfL_z}  \right\rfloor \in\bbZ^p,
\end{equation}
where $\sfL_z>0$ is a publicly known scale factor.
The state dynamics of each agent $i\in\calV$ is given by 
\begin{align}\label{eq:metroQuant}
     z_i(t+1)  = z_i(t) + \sfL_z  \sum_{j\in\calN_i}  w_{ij}(\sfQ(z_j(t)) -\sfQ(z_i(t))), 
\end{align}
where $z_i(0) = z_i^\ini\in\bbR^p$ is the initial value.

The diffusive coupling term $\sum_{j\in\calN_i}  w_{ij}(\sfQ(z_j(t)) -\sfQ(z_i(t)))$ enforces consensus of the states $z_i(t)$ provided that the weights $w_{ij}\in\bbR$ are chosen appropriately. 
Various methods have been proposed in \cite{XiaoBoyd04}, but we particularly adopt the \textit{Metropolis weights} as they only require local information:
\begin{align}\label{eq:wij}
    w_{ij} &:= \frac{1}{2\left(1+\max\{|\calN_i|, |\calN_j| \}\right) }, & \forall (i,j) &\in \calE.
\end{align}

The following lemma states that $z_i(t)$ \textit{practically converges} to $z^\avg:=\frac{1}{M}\sum_{l=1}^M z_l^\ini$ for all $i\in\calV$.
The convergence error can be made arbitrarily small by choosing a sufficiently small scale factor $\sfL_z$, as it reduces the precision losses introduced by the quantizer in \eqref{eq:Lz}.

\begin{lem}[{\cite[Theorem~1, Proposition~4]{ChamLiuj16}}]\label{lem:quantMetro}\upshape 
    Consider the dynamics \eqref{eq:metroQuant} with the weights \eqref{eq:wij}. 
    For any initial values $\{z_i^\ini\}_{i\in\calV}$, it holds that
    \begin{equation*}
        \limsup_{t\to\infty} \left\|z_i(t)- z^\avg \right\| \le \sfL_z, ~~~~ \forall i \in \calV.
    \end{equation*}
\end{lem}

To facilitate the use of \eqref{eq:wij}, we assume the following.
\begin{asm}\upshape\label{asm:metro}
    Each agent $i\in\calV$ knows the agents that are within two hops of itself. 
\end{asm}

This assumption is mild, since two-hop neighbor information can be obtained locally, e.g., by exchanging neighbor lists during initialization.
Moreover, similar assumptions have been commonly made in the resilient consensus literature \cite{GaorYang23,HadjDomi24,XiegShen25}, as it enables neighbor monitoring and attack detection.

\subsection{Protocol design}\label{subsec:avgProtocol}

Our objective is to design a secure implementation of the dynamics \eqref{eq:metroQuant} utilizing additive secret sharing, so that no individual agent can infer the contribution of other agents from the messages it receives. 

We first assume that all agents share another common scale factor $\sfL_w>0$ such that
\begin{equation}\label{eq:Lw}
    \bar{w}_{ij}:=\frac{w_{ij}}{\sfL_w} \in \bbZ, ~~~~ \forall (i,j)\in\calE.
\end{equation}
Such a scale factor always exists because the weights $w_{ij}$ in \eqref{eq:wij} are rational numbers.
For example, if the agents agree to use a fixed-point representation with three fractional bits, then choosing $\sfL_w=1/8$ ensures \eqref{eq:Lw}.

Then, we propose to modify the dynamics \eqref{eq:metroQuant} as 
\begin{align}\label{eq:metroFinal}
     &\bar{z}_i(t+1) = \bar{z}_i(t) \\
     &+\sfL_w\sfL_z\left( \phi_{ii}(t)  + \sum_{j\in\calN_i} \left(\zeta_{ij}(t) - \bar{w}_{ij}\sfQ(\bar{z}_i(t))\right) \modp q \right), \nonumber
\end{align}
where $\bar{z}_i(t)\in\bbR^p$ is the state with the initial value $\bar{z}_i(0)=z_i^\ini$, and $\zeta_{ij}(t)$ is the message received from agent $j\in\calN_i$, defined by\footnote{Indeed, agent $j$ can compute $\zeta_{ij}(t)$, as it knows $|\calN_i|$ under Assumption~\ref{asm:metro}, and can therefore compute its associated weight $\bar{w}_{ij}$.}
\begin{equation}\label{eq:theta}
    \zeta_{ij}(t) := \bar{w}_{ij}\sfQ(\bar{z}_j(t)) + \phi_{ij}(t) \modp q \in \bbZ_q^p.
\end{equation}
Here, the quantities $\phi_{ii}(t)\in\bbZ_q^p$ and $\phi_{ij}(t)\in\bbZ_q^p$ are masking terms, the former known only to
agent $i$ and the latter only to agent $j\in\calN_i$, that satisfy 
\begin{equation}\label{eq:zeroshare}
    \phi_{ii}(t) + \sum_{j\in\calN_i} \phi_{ij}(t) \modp q = \bfzero.
\end{equation}

Note that agent $i$ cannot recover  $\bar{w}_{ij}\sfQ(\bar{z}_j(t))$ from $\zeta_{ij}(t)$, since it does not have access to the masking term $\phi_{ij}(t)$. 
A method for generating the masking terms that satisfy \eqref{eq:zeroshare} while remaining known only to the corresponding agent will be presented in Section~\ref{subsec:mask}.
The entire procedure of the proposed secure distributed average consensus protocol is summarized in
Protocol~\ref{alg:metro}.

The following theorem states that the states $\bar{z}_i(t)$ practically converge to $z^\avg$, provided that the modulus $q$ is chosen sufficiently large for given $\sfL_z$ and $\sfL_w$.
A sufficiently large $q$ ensures that the modulo operation in \eqref{eq:metroFinal} does not cut-off the higher bits of the integer quantities.

To state the result, define the weight matrix $W=[W_{ij}]\in\bbR^{M\times M}$ by
\begin{align}\label{eq:W}
    W_{ij} := 
    \begin{cases}
        w_{ij}, & j\in\calN_i, \\
        1-\sum_{l\in\calN_i}w_{il}, & j=i,\\
        0, & \mbox{otherwise},
    \end{cases}
\end{align}
and introduce
\begin{align*}
    \lambda &:= \rho\left(W-\frac{1}{M}\bfone \bfone^\top \right), & \tilde{z}^\ini_{\max} &:= \max_{i\in\calV} \left\| z_i^\ini - z^\avg \right\|.
\end{align*}
Here, $\bfone\in\bbR^M$ is the vector of all ones and $\rho(\cdot)$ denotes the spectral radius.
In particular, $0 \le \lambda<1$ as shown in Appendix~\ref{apdx:tech}.
This quantity governs the rate at which disagreement among the agents decays, with a smaller $\lambda$ yielding faster convergence to consensus.

\begin{algorithm}[t]
    \caption{Secure distributed average consensus}\label{alg:metro}
    \begin{algorithmic}[1]
    \renewcommand{\algorithmicrequire}{\textbf{Input:}}
    \Require The graph $\calG$, scale factors $\sfL_z,\sfL_w>0$, number of iterations $T\in\bbN$, and initial values $\{z_i^\ini\}_{i\in\calV}$ 
    \renewcommand{\algorithmicrequire}{\textbf{Output:}}
    \Require $\bar{z}_i(T)$ for all $i\in\calV$
    \State Initialize $\bar{z}_i(0)=z_i^\ini$ for all $i\in\calV$
    \For{$t=0,1,\ldots,T-1$}
    \For{$i\in\calV$}  \label{line:start}
    \State Generate $\phi_{ii}(t)$ and $\{\phi_{ij}(t)\}_{j\in\calN_i}$ via \eqref{eqs:mask}
    \State Each agent $j\in\calN_i$ sends \eqref{eq:theta} to agent $i$
    \State Agent $i$ updates $\bar{z}_i(t+1)$ according to \eqref{eq:metroFinal}
    \EndFor\label{line:end}
    \EndFor
    \end{algorithmic}
\end{algorithm}

 \begin{thm}\upshape\label{thm:metroQuant2}
    Consider the dynamics \eqref{eq:metroFinal} subject to \eqref{eq:zeroshare}.
    Under Assumption~\ref{asm:metro}, for any initial values $\{z_i^\ini\}_{i\in\calV}$,
    if the parameters $\sfL_z>0$, $\sfL_w>0$, and $q\in\bbN$ satisfy
    \begin{align}\label{eq:qbound}
        q > \frac{M}{\sfL_w} \left( \frac{1}{2} + \frac{\|z^\avg\| + \tilde{z}_{\max}^\ini\sqrt{Mp} }{\sfL_z} + \frac{\sqrt{Mp}}{2(1-\lambda)}  \right),
    \end{align}
    then $z_i(t)\equiv \bar{z}_i(t)$ for all $i\in\calV$, and thus, 
    \begin{align}\label{eq:thmFinalMetroToShow}
        \limsup_{t\to\infty} \left\|\bar{z}_i(t)- z^\avg \right\| \le \sfL_z, ~~~~ \forall i \in \calV.
    \end{align}
\end{thm}
\noindent\textit{Proof.} See Appendix~\ref{apdx:metroQuant2}. \hfill $\blacksquare$

Given that the initial values of the agents are bounded by a known constant, \eqref{eq:qbound} is verifiable a priori, because $\{\lambda,M\}$ depend on the graph topology and $\{\sfL_z,\sfL_w\}$ are design parameters.
Moreover, we argue that overestimating $q$ is not critical since it affects neither the convergence error in \eqref{eq:thmFinalMetroToShow} nor the privacy guarantee, and increases the communication and computation costs only logarithmically, as each element of $\bbZ_q$ requires $\lceil \log_2 q \rceil$ bits.

We provide a guideline for choosing the parameters $\{\sfL_z, \sfL_w, q\}$. 
In practice, we recommend choosing $\sfL_z$ first, considering the trade-off that decreasing $\sfL_z$ reduces the convergence error in \eqref{eq:thmFinalMetroToShow}, but simultaneously increases the resolution of the quantizer which may be costly in implementation.
Next, choose $\sfL_w$ as large as possible subject to \eqref{eq:Lw}. 
This is desirable since it does not affect the convergence error, while a larger $\sfL_w$ decreases the lower bound on $q$ in \eqref{eq:qbound}.
Finally, select $q$ as small as possible while satisfying \eqref{eq:qbound}, since a larger modulus may increase computation and communication costs.

% The boundedness condition \eqref{eq:metroQuant2Cond} ensures that the higher bits of the summation term are not cut off by the modulo operation.
% This can be readily met by choosing a sufficiently large modulus $q$ because the summation term remains bounded for all $t\ge0$; the boundedness follows from the practical convergence of the states $z_i(t)$ and the finite network size of $\calG$, which imply the boundedness of $\sfQ(z_j(t)) - \sfQ(z_i(t))$ and $\bar{w}_{ij}$, respectively.

\subsection{Generation of masking terms}\label{subsec:mask}
We adopt the fully distributed algorithm presented in \cite{AlexDaru19} for generating the masking terms $\phi_{ii}(t)$ and $\{\phi_{ij}(t)\}_{j\in\calN_i}$ satisfying \eqref{eq:zeroshare}. 
For notational simplicity, we define 
$\calN_l^+ := \calN_l \cup \{l\}$ for all $l\in\calV$. 
The following assumption on the graph topology is imposed, as in \cite{AlexDaru19}. 
\begin{asm}\upshape\label{asm:graph}
    For every edge $(i,j)\in\calE$, the set $\calN_i\cap\calN_j $ is nonempty, i.e., $i$ and $j$ share at least one common neighbor. 
\end{asm}

This assumption essentially strengthens the graph's local connectivity.
It requires every edge to lie in a triangle and it is satisfied by densely connected or complete graphs.
Similar conditions have been frequently assumed in the literature on resilience and fault tolerance of distributed algorithms; see, e.g., \cite{SeneSund19,YuanIshi21,YuanIshi24,CaomMors08}.

\begin{figure}[t]
    \centering
    \includegraphics[width=0.98\linewidth]{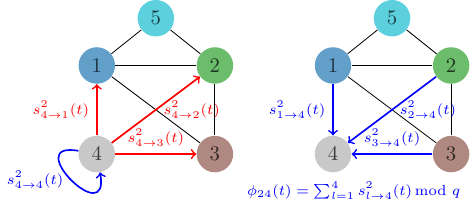}
    \caption{Illustration of $\calG$ with $5$ agents. 
    The left and right diagrams depict \eqref{eq:jshare} and \eqref{eq:mask}, respectively, for $i=2$ and $j=4$. In this case, $\calN_i^+\cap \calN_j^+=\{1,2,3,4\}$.
    % Left: Agent $j$ generating $n_{ij}$-shares of zero and distributing them as in \eqref{eq:jshare}. Right: Agent $j$ computing the masking part $\phi_{ij}(t)$ as in \eqref{eq:mask}.
    }
    \label{fig:graph}
\end{figure}

At each time step $t\ge 0$ and for a fixed aggregating agent $i$, let each neighbor $j\in\calN_i$ generate $|\calN_i^+\cap\calN_j^+|$-shares of the $p$-dimensional zero vector as 
\begin{subequations}\label{eqs:mask}
\begin{equation}\label{eq:jshare}
     (s^i_{j\to l}(t))_{l\in \calN_i^+\cap\calN_j^+} \leftarrow \Share(\bfzero,|\calN_i^+\cap\calN_j^+|), 
\end{equation}
and send each share $s^i_{j\to l}(t)\in\bbZ_q^p$ to the corresponding agent $l$.
Similarly, let the aggregating agent $i$ generate 
\begin{equation}\label{eq:ishare}
     (s^i_{i\to l}(t))_{l\in \calN_i^+} \leftarrow \Share(\bfzero,|\calN_i^+|), 
\end{equation}
and send $s^i_{i\to l}(t)\in\bbZ_q^p$ to the corresponding agent $l$.
Note that one of the generated shares is retained by the generating agent, since $j\in\calN_i^+ \cap \calN_j^+$ and $i\in\calN_i^+$.

Subsequently, each neighbor $j\in\calN_i$ and the aggregating agent $i$ take the summation of their own share and the shares it received from others, as
\begin{equation}\label{eq:mask}
    \begin{split}
        \phi_{ij}(t) &= \sum_{l\in\calN_i^+\cap\calN_j^+}  s^i_{l\to j}(t)  \modp q, ~~~~ \forall j\in\calN_i, \\
        \phi_{ii}(t) &= \sum_{l\in\calN_i^+}  s^i_{l\to i}(t) \modp q.
    \end{split}
\end{equation}
\end{subequations}
This procedure is illustrated in Fig.~\ref{fig:graph}.

\begin{lem}\upshape\label{lem:zeroshare}
    For every $i\in\calV$, the masking terms $\phi_{ii}(t)$ and $\{\phi_{ij}(t)\}_{j\in\calN_i}$ generated according to \eqref{eqs:mask} satisfy \eqref{eq:zeroshare}.
\end{lem}

\noindent\textit{Proof.} 
It directly follows from \eqref{eq:mask} that  
\begin{align*}
    \phi_{ii}(t) \!+\! \sum_{j\in\calN_i} \phi_{ij}(t) \modp q &= \sum_{j\in\calN_i^+} \sum_{l\in\calN_i^+\cap\calN_j^+} \!\!s^i_{l\to j}(t) \modp q  \\
    &= \sum_{j\in\calN_i^+} \sum_{l\in\calN_i^+\cap\calN_j^+}  \!\!s^i_{j\to l}(t) \modp q, 
\end{align*}
where the second equality holds because $j\in\calN_i^+$ and $l\in\calN_i^+\cap\calN_j^+$ if and only if $l\in\calN_i^+$ and $j\in\calN_i^+\cap\calN_l^+$.
Since
\begin{equation*}
    \sum_{l\in \calN_i^+\cap\calN_j^+} s^i_{j\to l}(t) \modp q = \bfzero, ~~~~ \forall j\in\calN_i^+,
\end{equation*}
due to \eqref{eq:jshare} and \eqref{eq:ishare}, this concludes the proof. \hfill $\blacksquare$

We briefly comment on the role of Assumption~\ref{asm:graph}.
Suppose it fails, i.e., agents $i\in\calV$ and $j\in\calN_i$ have no common neighbor, so that $\calN_i^+\cap\calN_j^+ = \{i,j\}$. 
Then, it follows from \eqref{eq:jshare} and \eqref{eq:mask} that
\begin{align*}
    s_{j\to i}^i(t) + s_{j\to j}^i(t) \modp q &= \bfzero, \\
    s_{i\to j}^i(t) + s_{j\to j}^i(t) \modp q &= \phi_{ij}(t).
\end{align*}
Since agent $i$ knows both $s_{j\to i}^i(t)$ and $s_{i\to j}^i(t)$, it can reconstruct
$s_{j\to j}^i(t)$ and $\phi_{ij}(t)$, thus compromising privacy.
Assumption~\ref{asm:graph} rules this out because it guarantees that $|\calN_i^+\cap\calN_j^+|\ge 3$, so that no individual agent can reconstruct another agent's masking term.

The communication complexity of the proposed protocol is also analyzed. 
In \eqref{eq:metroFinal}, each agent transmits its state to each neighbor per iteration, resulting in a total of $|\calE|$ messages.
Our protocol additionally distributes the shares to construct the masking terms as in \eqref{eq:mask}, requiring
\begin{align}\label{eq:commComplex}
    \sum_{i\in\calV} |\calN_i| + \sum_{i\in\calV}\sum_{j\in\calN_i} |\calN_i^+ \cap \calN_j| \le |\calE|  + \Delta|\calE|,
\end{align}
messages to be exchanged
per iteration, where $\Delta := \max_{i\in\calV} |\calN_i|$.
Therefore, the communication complexity of the proposed protocol scales linearly with that of the standard average consensus.

\begin{rem}\upshape
Using the same masking generation procedure as in \eqref{eqs:mask}, \cite{AlexDaru19} also proposed a secure average consensus protocol; however, our protocol improves upon their approach in two key aspects.
\begin{enumerate}[leftmargin=*]
    \item In \cite{AlexDaru19}, the weights $w_{ij}$ and the states $z_i(t)$ in \eqref{eq:metroQuant} are assumed to take values in $\bbZ_q$, and the effect of quantization is therefore not addressed. 
    The proposed protocol accommodates real-valued weights and states through the quantizer in \eqref{eq:Lz}. 
    The effect of quantization on convergence error is analyzed in Theorem~\ref{thm:metroQuant2}, which provides an explicit guideline for choosing the scale factors $\sfL_w$, $\sfL_z$, and the modulus $q$.
    \item In \cite{AlexDaru19}, the weights $w_{ij}$ are encrypted and distributed to the agents by a trusted central server.
    By specifically adopting the Metropolis weights, our design allows each agent to locally compute its associated weights under Assumption~\ref{asm:metro}---the same assumption used in \cite{AlexDaru19}.
    Thus, our method facilitates a fully distributed design and eliminates the need for encryption, without imposing any additional assumptions.   
\end{enumerate}
 
\end{rem}

\begin{rem}
The present analysis assumes that all agents remain active throughout the protocol, no packet losses occur during transmission, and the communication is synchronous.
Nevertheless, the protocol can tolerate agent dropouts after completing Lines~\ref{line:start}--\ref{line:end} of Protocol~\ref{alg:metro} at each iteration, as the masking terms are generated independently at each time step and are not reused.
Asynchronous communication is also admissible provided that all messages are exchanged reliably, as it does not affect correctness, but at the cost of an increased runtime due to communication delays.
In contrast, dropout during masking term generation leads to the failure of the protocol, since \eqref{eq:zeroshare} no longer holds.  
Handling such dropout requires additional mechanisms and is left for future work.
\end{rem}

\subsection{Privacy analysis}\label{subsec:priv}

We adopt the paradigm of simulation based security \cite{Gold01} and formally establish the security of Protocol~\ref{alg:metro}. 
Let $\sfx_i\in\{0,1\}^*$ denote the \textit{private input} of agent $i\in\calV$, where $\{0,1\}^*$ denotes the set of binary strings of arbitrary length.
Suppose that the objective of the agents is to design a protocol $\Pi$ that realizes a desired functionality $\calF=(\calF_1,\ldots,\calF_M)$, where each $\calF_i: \{0,1\}^* \times \cdots \times \{0,1\}^* \to \{0,1\}^*$ computes a \textit{private output}
\begin{equation}\label{eq:fi}
    \sfy_i = \calF_i(\sfx_1,\ldots,\sfx_M),
\end{equation}
for agent $i$.

Ideally, this can be achieved by submitting $(\sfx_1,\ldots,\sfx_M)$ to a trusted third party, which then computes and securely returns $\sfy_i$ to the corresponding agent. 
In practice, however, such a trusted third party is typically unavailable. 
Therefore, the agents must jointly evaluate the functionality through a distributed protocol.
This introduces the possibility of agents learning additional information beyond their private input and output from the messages exchanged during the execution of the protocol.

In this context, we say that a protocol $h$-privately computes a functionality in the presence of semi-honest agents if any coalition $\calC\subset \calV$ with $|\calC|\le h$ learns nothing beyond its own private input/output pairs $(\sfx_i,\sfy_i)_{i\in\calC}$. 
The semi-honest adversarial model is a standard starting point, since protocols developed under this model can often be strengthened to covert or malicious settings using standard cryptographic techniques \cite{HemeWels14}. 
For example, generating the shares based on verifiable secret sharing \cite{ChorGold85,Feld87} could allow each agent to check whether the received shares satisfy \eqref{eq:zeroshare}.
It is also well motivated in our setting, as the agents are primarily interested in obtaining the correct output rather than disrupting the protocol.
This notion is formalized through the following definition.

\begin{defn}\upshape\label{defn:security}
    Let $\msg_i$ denote the collection of all messages received by agent $i\in\calV$ during the execution of a deterministic protocol $\Pi$ on a security parameter $\kappa\in\bbN$.
    The protocol $h$-privately computes a functionality $\calF$ in the presence of semi-honest agents if there exists a probabilistic polynomial-time simulator\footnote{ A randomized algorithm whose running time is bounded by a fixed polynomial in $\kappa$.} $\Sim_\calC$ for each coalition $\calC\subset \calV$ with $|\calC|\le h$ such that
    \begin{align}\label{eq:privGoal}
        (\sfx_i,\msg_i)_{i\in \calC}\equiv \Sim_\calC \left( (\sfx_i,\sfy_i)_{i\in\calC} \right),
    \end{align}
    where $\equiv$ denotes statistical indistinguishability.
\end{defn}

The existence of a simulator implies that any coalition $\calC$ of bounded size $h$ cannot distinguish whether it is participating in the actual protocol $\Pi$ or receiving virtual messages generated by $\Sim_\calC$.
Since the simulator only has access to the coalition's private input/output pairs $(\sfx_i,\sfy_i)_{i\in\calC}$, any information contained in $(\msg_i)_{i\in\calC}$ can be reconstructed from these pairs, and thus, no additional information of the non-colluding agents is revealed.

To apply this definition to  Protocol~\ref{alg:metro}, we specify the private input and output of each agent $i$ as
\begin{align}\label{eq:privIn}
    \sfx_i=(z_i^\ini,\{\bar{w}_{ij},\bar{w}_{ji}\}_{j\in\calN_i}), ~~~~\sfy_i=(\bar{z}_i(t))_{t=1}^T.
\end{align}
The input $\sfx_i$ consists of all information locally available to agent $i$ that is used internally. 
The output $\sfy_i$ is the sequence of local state updates that the agent obtains.
The following theorem establishes the security of Protocol~\ref{alg:metro}.

\begin{thm}\upshape\label{thm:priv}
    Under Assumptions~\ref{asm:metro} and~\ref{asm:graph}, Protocol~\ref{alg:metro} $h$-privately computes the functionality $\calF=(\calF_1,\ldots,\calF_M)$ of the form \eqref{eq:fi} with respect to \eqref{eq:privIn} in the presence of semi-honest agents, where $h=\min_{(i,j)\in\calE}|\calN_i^+ \cap \calN_j^+|-2$. 
\end{thm}

\noindent\textit{Proof.} See Appendix~\ref{apdx:priv}. \hfill $\blacksquare$

\begin{algorithm}[t]
    \caption{Privacy-preserving fully distributed GPR}\label{alg:poe}
    \begin{algorithmic}[1]
    \renewcommand{\algorithmicrequire}{\textbf{Input:}}
    \Require The graph $\calG$, scale factors $\sfL_z,\sfL_w>0$, number of iterations $T\in\bbN$, and a test point $x\in\bbR^n$
    \renewcommand{\algorithmicrequire}{\textbf{Output:}}
    \Require $\hat{f}_i^{(T)}(x)$ and $V_i^{(T)}(x)$ for all $i\in\calV$
    \State Each agent $i\in\calV$ computes $z_i^\ini$ as in \eqref{eq:init}
    \State Invoke Protocol~\ref{alg:metro} with $z_i^\ini$ and $T$
    \State Define $[\bar{z}_{i,1}(T);\bar{z}_{i,2}(T)] := \bar{z}_i(T)$
    \State $V^{(T)}_i(x) \gets 1/\bar{z}_{i,2}(T)$
    \State $\hat{f}^{(T)}_i(x) \gets V^{(T)}_i(x) \cdot \bar{z}_{i,1}(T)$
    \end{algorithmic}
\end{algorithm}

Theorem~\ref{thm:priv} guarantees that each agent’s privacy is preserved against any semi-honest coalition of size at most $h$.
Since $h$ increases with the minimum number of common neighbors between neighboring agents, the proposed protocol becomes more robust to collusion as the graph $\calG$ becomes more densely connected.
In practice, $h$ can be increased either by adding more connections or by introducing ``dummy'' agents that participate only in the masking generation procedure but do not contribute to the consensus protocol itself.
However, it should be noted that these approaches come at the cost of higher communication overhead, as reflected in \eqref{eq:commComplex}.

\section{Privacy-preserving Fully Distributed GPR}\label{sec:dgpr}

% We present the proposed \textit{privacy-preserving fully distributed GPR} protocol that allows the agents to jointly compute an approximation of \eqref{eq:DGPR}, while preserving the confidentiality of their state information from neighboring agents.
% We analyze the convergence of the proposed protocol, and provide a method to optimize the kernel hyperparameters in a privacy-preserving and fully distributed manner, by integrating Protocol~\ref{alg:metro} into a consensus based gradient method.

\subsection{Protocol design}

The proposed \textit{privacy-preserving fully distributed GPR} protocol is summarized in Protocol~\ref{alg:poe}.
The main idea is to invoke Protocol~\ref{alg:metro} to collaboratively compute the summations $\sum_{i=1}^M V_i^{-1}(x)\cdot \hat{f}_i(x)$ and $\sum_{i=1}^M V_i^{-1}(x)$ in \eqref{eq:DGPR}. 
By the security of Protocol~\ref{alg:metro}, each agent cannot learn individual contributions of its neighbors, beyond what is already implied by the aggregated summations.
Each agent then locally inverts the latter to compute $V(x)$, and multiplies it with the former to compute $\hat{f}(x)$.

The protocol is described in detail below.
During the offline procedure, each agent $i\in\calV$ computes \eqref{eq:posterior} for a given test point $x\in\bbR^n$, and defines the initial value $z_i^\ini\in\bbR^2$ as 
\begin{equation}\label{eq:init}
    z_i^\ini = 
    M\cdot
    \begin{bmatrix}
        V_i^{-1}(x)\cdot\hat{f}_i(x)  \\
        V_i^{-1}(x)
    \end{bmatrix}.
\end{equation}
Here, we assume that the agents know the network size $M$, as it can be obtained by running a distributed network size estimation algorithm, e.g., \cite{LeedLees18}, a priori.
Such estimation does not compromise privacy, as it is independent of the agents' local datasets.

During the online procedure, we invoke Protocol~\ref{alg:metro} with respect to \eqref{eq:init}.
Then, with the output $\bar{z}_i(T)=:[\bar{z}_{i,1}(T);\bar{z}_{i,2}(T)]\in\bbR^2$, each agent $i\in\calV$ computes 
\begin{align}\label{eq:privPoe}
    V^{(T)}_i(x) := \frac{1}{\bar{z}_{i,2}(T)}, ~~~~ \hat{f}^{(T)}_i(x) := V^{(T)}_i(x) \cdot \bar{z}_{i,1}(T).
\end{align}

The following theorem states that the outputs $\hat{f}^{(T)}_i(x)$ and $V^{(T)}_i(x)$ of Protocol~\ref{alg:poe} practically converge to the outputs $\hat{f}(x)$ and $V(x)$ of standard distributed GPR, respectively, where the convergence error can be made arbitrarily small by decreasing the scale factor $\sfL_z$ and increasing $T$.

\begin{figure*}[t]
    \begin{centering}
        \includegraphics[width=0.98\linewidth]{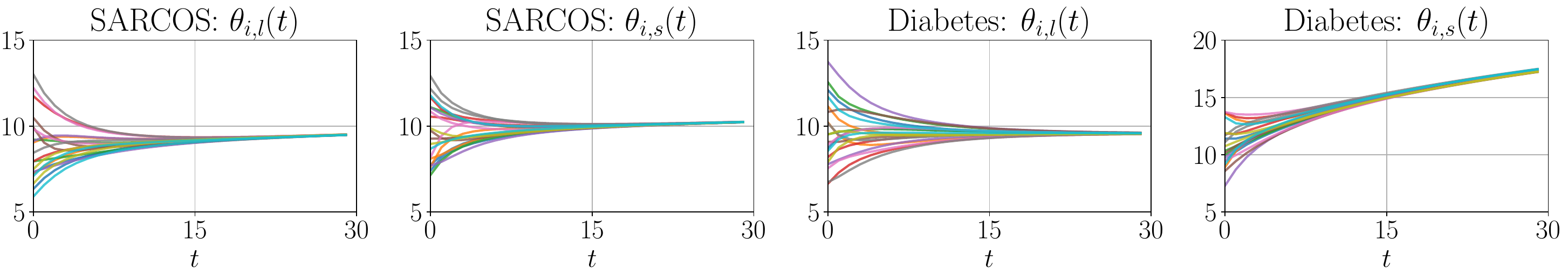}
        \caption{Convergence of the hyperparameter estimates $\theta_{i,l}(t)$ and $\theta_{i,s}(t)$ across $20$ agents when using \eqref{eqs:hyper}. 
        For the SARCOS dataset, we only show the estimates corresponding to the first output dimension as a representative case.} 
        \label{fig:hypBenchmark}
    \end{centering}
\end{figure*}

\begin{thm}\label{thm:poe}\upshape
    Given the parameters $\sfL_z>0$, $\sfL_w>0$, and $q\in\bbN$, and a test point $x\in\bbR^n$, assume that 
    \begin{equation}\label{eq:xi}
        \frac{1}{V(x)} = \sum_{i=1}^M V_i^{-1}(x) > \sfL_z .
    \end{equation}
    If \eqref{eq:qbound} holds, then there exist class $\calK$ functions $\beta:\bbR_{\ge 0} \to \bbR_{\ge 0}$ and $\gamma:\bbR_{\ge 0}\to \bbR_{\ge 0}$ such that 
    \begin{subequations}\label{eq:thmPoeToShow}
    \begin{align}
        \limsup_{T\to\infty} \left\| \hat{f}^{(T)}_i(x) -\hat{f}(x) \right\|  &\le \beta(\sfL_z), \label{eq:thmPoeToShow1}\\
        \limsup_{T\to\infty} \left\| V^{(T)}_i(x) -V(x) \right\| &\le \gamma(\sfL_z), \label{eq:thmPoeToShow2} 
    \end{align}
    \end{subequations}
    for all $i\in\calV$.
\end{thm}

\noindent \textit{Proof.} See Appendix~\ref{apdx:poe}, where the functions $\beta$ and $\gamma$ are explicitly constructed. \hfill $\blacksquare$

We argue that the condition \eqref{eq:xi} is rather mild.
Indeed, it follows from \eqref{eq:posterior} and the positive definiteness of $k$ that $0<V_i(x)\le k(x,x)$, and thus, $\sfL_z<M/k(x,x)$ is sufficient for \eqref{eq:xi} to hold.
This can be verified a priori, as it depends only on the network topology and the kernel, and can be easily satisfied by choosing $\sfL_z$ sufficiently small.

The security of Protocol~\ref{alg:poe} directly follows from that of Protocol~\ref{alg:metro}.

\begin{cor}\upshape
    Fix a test point $x\in\bbR^n$, and let $\sfx_i=(\calD_i,\{\bar{w}_{ij},\bar{w}_{ji}\}_{j\in\calN_i})$ and $\sfy_i = (\hat{f}_i^{(T)}(x), V_i^{(T)}(x))$ for each $i\in\calV$. 
    Under Assumptions~\ref{asm:metro} and~\ref{asm:graph}, Protocol~\ref{alg:poe} $h$-privately computes the functionality $\calF=(\calF_1,\ldots,\calF_M)$ of the form \eqref{eq:fi} in the presence of semi-honest agents, where $h=\min_{(i,j)\in\calE}|\calN_i^+ \cap \calN_j^+|-2$.
\end{cor}

\noindent \textit{Proof.}
    Protocol~\ref{alg:poe} augments Protocol~\ref{alg:metro} with a pre-processing step of computing $z_i^\ini$ from $\calD_i$ via \eqref{eq:init}, and a post-processing step of computing $\sfy_i$ from $\bar{z}_i(T)$ via \eqref{eq:privPoe}.
    Neither step involves communication with the neighboring agents, so the security of Protocol~\ref{alg:poe} directly follows from Theorem~\ref{thm:priv}. 
    This concludes the proof. \hfill $\blacksquare$

Consequently, no semi-honest coalition $\calC\subset\calV $ of size at most $h$ learns any information beyond what is already implied by the local datasets $(\calD_i)_{i\in\calC}$ and by the outputs $(\hat{f}_i^{(T)}(x),V_i^{(T)}(x))_{i\in\calC}$ when running Protocol~\ref{alg:poe}.

% \begin{figure*}[t]
%     \begin{centering}
%         \includegraphics[width=0.98\linewidth]{AAAI/img/hyp.pdf}
%         \caption{Results of the privacy-preserving fully distributed hyperparameter optimization procedure described by \eqref{eqs:hyper}. 
%         (a,b) Practical convergence of the estimates $\theta_{i,l}(t)$ and $\theta_{i,s}(t)$.
%         (c) Monotonic increase of $\widehat{\mathsf{LML}}(t)$ over iteration.} 
%         \label{fig:hyperparameter}
%     \end{centering}
% \end{figure*}
% \begin{figure*}[t]
%     \begin{centering}
%         \includegraphics[width=0.98\linewidth]{AAAI/img/consensus.pdf}
%         \caption{Results of the Protocol~\ref{alg:metro}. 
%         (a) Local dataset $\calD_i$ of each agent $i\in\calV$. 
%         (b) Local estimates $\hat{f}_i(x)$ (solid line) with $2\sqrt{V_i(x)}$ confidence intervals (shaded area). 
%         (c) Final estimates  $\hat{f}_i^{(T)}$ after $T$ iterations (solid line) with $2\sqrt{V_i^{(T)}(x)}$ confidence intervals (shaded area).} 
%         \label{fig:consensus}
%     \end{centering}
% \end{figure*}

\subsection{Hyperparameter optimization}\label{subsec:hyp}

A widely used positive definite kernel is the squared-exponential kernel, defined by
\begin{equation*}\label{eq:kernel}
    k(x,x') = \theta_s^2 \cdot \exp \left( -\frac{1}{2} \frac{\|x-x' \|_2^2}{\theta_l^2}\right),
\end{equation*}
where the hyperparameters $\theta_l>0$ and $\theta_s>0$ determine the length-scale and signal variance of $k$, respectively.
It is well recognized that the choice of kernel hyperparameters significantly affects the prediction accuracy of GPR \cite{WillRasm06}.
Accordingly, we develop a method for optimizing the kernel hyperparameters in a privacy-preserving manner by exploiting Protocol~\ref{alg:metro}.
The squared-exponential kernel is adopted solely for the simplicity of presentation, and does not restrict the generality of the proposed method.

A common approach for hyperparameter optimization is to maximize the log marginal likelihood of the entire dataset $\calD:=\{\calD_i\}_{i=1}^M$ conditioned on the hyperparameters.
However, this is not applicable in our setting, as the local datasets are considered private.
Therefore, following \cite{DeisNgju15}, we neglect cross covariances between the local datasets and approximate the log marginal likelihood as the sum of the log marginal likelihoods of each dataset, i.e., 
\begin{equation}\label{eq:lml}
    \log p(\calD \mid \Theta) \approx \sum_{i=1}^M \log p(\calD_i\mid \Theta)=: \sum_{i=1}^M h_i(\Theta),
\end{equation}
where $\Theta:=[\theta_l;\theta_s]\in\bbR^2$. 
Each summand in \eqref{eq:lml} is locally computable by the corresponding agent, as 
\begin{align*}
    h_i(\Theta) &= -\frac{\bfy_i^\top \left(\bfK_i+\sigma_\epsilon^2\cdot I_{N_i} \right)^{-1} \bfy_i}{2} \\
    & -\frac{\log \det(\bfK_i+\sigma_\epsilon^2\cdot I_{N_i})}{2} - \frac{N_i}{2} \log 2\pi.
\end{align*}

Note that the gradient of the right-hand-side of \eqref{eq:lml} is given by $\sum_{i=1}^M \nabla h_i(\Theta)$, where each term can be locally computed by the agents.
We therefore propose that the agents alternate between invoking Protocol~\ref{alg:metro} to reach a consensus in a privacy-preserving manner and performing a local gradient ascent step to maximize their respective log marginal likelihoods.
This alternating strategy is commonly known in the literature as the \textit{consensus based gradient method} \cite{NediOlsh08}.

The overall procedure is as follows: 
Each agent $i\in\calV$ maintains an estimate of the optimal hyperparameters, denoted by $\Theta_i(t)=[\theta_{i,l}(t);\theta_{i,s}(t)]\in\bbR^2$.
At each time step $t\ge 0$, Protocol~\ref{alg:metro} is invoked with the initial value $z_i^\ini = \Theta_i(t)$ and $T=1$.
The output is used to update the local estimate, as 
\begin{subequations}\label{eqs:hyper}
\begin{equation}
    \Theta_i(t+1/2) = \bar{z}_i(T).
\end{equation}
Subsequently, each agent $i$ performs a local gradient ascent step, as 
\begin{equation}\label{eq:intUpdate}
    \Theta_i(t+1) = \Theta_i(t+1/2) + \eta(t) \nabla h_i(\Theta_i(t)),
\end{equation}
where $\eta(t)>0$ is the step size.
\end{subequations}

\begin{prop}\label{prop:hyper}
    Consider the dynamics \eqref{eqs:hyper} under Assumption~\ref{asm:metro}. 
    Assume that there exists a compact convex set $\Xi\subset\bbR_{>0}^2$ such that $\Theta_i(t)\in\Xi$ for all $i\in\calV$
and all $t\ge 0$. Define $r_\Theta := \max_{\Theta\in\Xi}\|\Theta\|$.
If the modulus $q$ satisfies
\begin{equation}\label{eq:hypParamsCond}
    q>\frac{M}{\sfL_w} \!\left(\frac{1}{2}+\frac{(1+2\sqrt{2M})r_\Theta}{\sfL_z} + \frac{\sqrt{2M}}{2(1-\lambda)}\right),
\end{equation}
then
\begin{multline}\label{eq:propHyperToShow}
    \Theta_i(t+1) = \Theta_i(t) + \eta(t) \nabla h_i(\Theta_i(t)) \\ 
    + \sfL_z \sum_{j\in\calN_i} w_{ij} \left( \sfQ(\Theta_j(t)) - \sfQ(\Theta_i(t)) \right),
\end{multline}
for all $i\in\calV$ and $t\ge 0$.
\end{prop}

\noindent \textit{Proof.}
At each time step $t\ge 0$, Protocol~\ref{alg:metro} is invoked with $z_i^\ini=\Theta_i(t)$, $p=2$, $T=1$, and $z^\avg = \frac{1}{M}\sum_{i=1}^M \Theta_i(t)$. Since $\Xi$ is convex, $z^\avg\in\Xi$, and hence, $\|z^\avg\|\le r_\Theta$ and $\tilde{z}_{\max}^\ini\le 2r_\Theta$.
Therefore, it follows from Theorem~\ref{thm:metroQuant2} that 
\begin{align*}
    \Theta_i(t+1/2) = \Theta_i(t) + \sfL_z \sum_{j\in\calN_i} w_{ij} \left( \sfQ(\Theta_j(t)) - \sfQ(\Theta_i(t)) \right),
\end{align*}
under \eqref{eq:hypParamsCond}. 
Then, \eqref{eq:propHyperToShow} follows from \eqref{eq:intUpdate}, and this concludes the proof.
\hfill $\blacksquare$

Proposition~\ref{prop:hyper} establishes that \eqref{eqs:hyper} reduces to the quantized consensus based gradient algorithm \eqref{eq:propHyperToShow} under \eqref{eq:hypParamsCond}.   
Analyzing its convergence is nevertheless nontrivial because $h_i$ is generally nonconcave and quantization introduces error terms. 
Therefore, only convergence to a neighborhood of a stationary point can be expected, and such a result is expected to follow by extending \cite{NediOlsh08,YoshHaya23} to the nonconcave setting.
However, this is beyond the scope of this work, and the proposed scheme is assessed empirically in Section~\ref{sec:simul}.

\section{Experiments and Results}\label{sec:simul}

This section provides experimental results to demonstrate the effectiveness of the proposed privacy-preserving fully distributed GPR protocol.\footnote{Code fully available at https://github.com/yj-jang-98/PP-dGPR.} 
We applied the protocol to two real-world benchmark datasets: the \textit{SARCOS} dataset \cite{WillRasm06}, which is widely used for evaluating the performance of GPR in large-scale problems, and the \textit{Diabetes} patient dataset \cite{EfroHast04}, where privacy concerns are particularly relevant.
From the SARCOS dataset, we randomly chose $4{,}000$ training samples and $200$ test samples, with $21$ input and $7$ output features.
The Diabetes dataset contains $442$ samples, with $10$ input features and a single output, from which roughly $20\%$ (89 samples) were randomly selected as test samples. 
For the multi-output SARCOS dataset, we trained a separate single-output model for each output dimension.
In this case, we interpret \eqref{eq:DGPR} and \eqref{eq:privPoe} as vector-valued functions, constructed by stacking the outputs of the $7$ trained GP models.
All experiments were conducted using Python $3.12.13$ and the GPyTorch $1.15.2$ library on Google Colab with an NVIDIA T4 GPU.

\subsection{Evaluation of hyperparameter optimization}

Fig.~\ref{fig:hypBenchmark} illustrates the results of the hyperparameter optimization procedure described in \eqref{eqs:hyper}.
The experiments were conducted on a network of $M=20$ agents, each connected to four neighbors, with training samples approximately evenly distributed across the agents.
The parameters were set to $\sfL_z=1/2^{20}$ and $\sfL_w=1/40$, and the initial hyperparameter estimates $\Theta_i(0)$ were sampled uniformly at random from the range $[5,15]^2$.
The right-hand-side of \eqref{eq:qbound} is approximately $2^{37}$ for this configuration, so we chose $q=2^{40}$.
The optimization was performed over $30$ iterations, with the decaying step size $\eta(t)=0.1\cdot 0.99^t$. 
As shown in Fig.~\ref{fig:hypBenchmark}, the local estimates $\Theta_i(t)$ converge to a common value across all agents.

\begin{figure*}[t]
    \begin{centering}
        \includegraphics[width=0.98\linewidth]{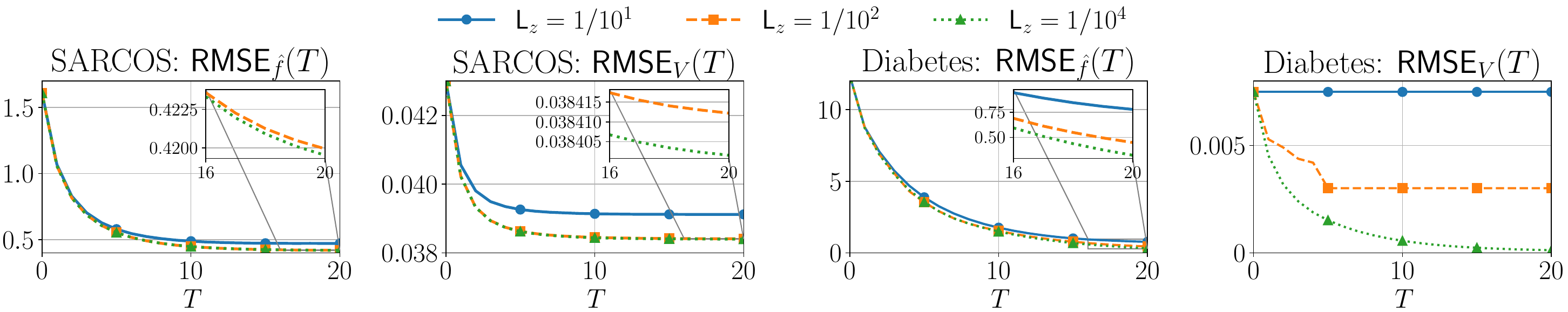}
        \caption{Monotonic decrease of RMSE as the number of iterations $T$ increases and the scale factor $\sfL_z$ decreases. } 
        \label{fig:accuracy}
    \end{centering}
\end{figure*}

\sisetup{uncertainty-mode=separate, table-number-alignment=center}
\begin{table*}[htbp]
    \centering
    \caption{ Comparison of total runtime and RMSE, where ``Degree'' is the number of neighbors per agent.}
    \label{tab:computation_time}
    \footnotesize
    \setlength{\tabcolsep}{2pt}

    \begin{tabular}{
        @{}
        c
        c
        c
        @{\hspace{5pt}}
        S[table-format=1.4]
        S[table-format=1.4]
        S[table-format=1.4]
        @{\hspace{5pt}}
        S[table-format=1.4]
        S[table-format=1.4]
        S[table-format=1.4]
        @{\hspace{5pt}}
        S[table-format=1.4(4)]
        S[table-format=1.4(4)]
        S[table-format=2.4(4)]
        S[table-format=1.4(4)]
        @{}
    }
    
    \toprule
    
    \multirow{2}{*}{\textbf{Dataset}}
        & \multirow{2}{*}{$M$}
        & \multirow{2}{*}{\textbf{Degree}}
        & \multicolumn{3}{c}{$\RMSE_{\smash{\hat{f}}}(T)$}
        & \multicolumn{3}{c}{$\RMSE_V(T)$}
        & \multicolumn{4}{c}{\textbf{Total runtime [sec] (mean $\pm$ std.)}} \\
        
    \cmidrule(r){4-6}
    \cmidrule(r){7-9}
    \cmidrule{10-13}
    
        & &
        & \multicolumn{1}{c}{\cite{LuojZhan23}}
        & \multicolumn{1}{c}{{\cite{NawaChen24}}}
        & \multicolumn{1}{c}{\textbf{Ours}}
        & \multicolumn{1}{c}{{\cite{LuojZhan23}}}
        & \multicolumn{1}{c}{{\cite{NawaChen24}}}
        & \multicolumn{1}{c}{\textbf{Ours}}
        & \multicolumn{1}{c}{\textbf{Distributed GPR}}
        & \multicolumn{1}{c}{{\cite{LuojZhan23}}}
        & \multicolumn{1}{c}{{\cite{NawaChen24}}}
        & \multicolumn{1}{c}{\textbf{Ours}} \\
        
    \midrule

    \multirow{3}{*}{SARCOS}
        & $20$ & $4$
        & \multicolumn{1}{c}{$*$}
        & \multicolumn{1}{c}{$<10^{-4}$}
        & 0.0192
        & \multicolumn{1}{c}{$*$}
        & \multicolumn{1}{c}{$<10^{-4}$}
        & 0.0003
        & 0.4797(1084)
        & \multicolumn{1}{c}{$>300$}
        & 7.2512(8093)
        & 0.5884(1112) \\

        & $20$ & $19$
        & \multicolumn{1}{c}{$*$}
        & \multicolumn{1}{c}{$<10^{-4}$}
        & 0.0004
        & \multicolumn{1}{c}{$*$}
        & \multicolumn{1}{c}{$<10^{-4}$}
        & \multicolumn{1}{c}{$<10^{-4}$}
        & 0.4631(1048)
        & \multicolumn{1}{c}{$>300$}
        & 7.2512(8093)
        & 0.9377(1073) \\

        & $40$ & $4$
        & \multicolumn{1}{c}{$*$}
        & \multicolumn{1}{c}{$<10^{-4}$}
        & 0.6767
        & \multicolumn{1}{c}{$*$}
        & \multicolumn{1}{c}{$<10^{-4}$}
        & 0.0113
        & 0.8686(1950)
        & \multicolumn{1}{c}{$>300$}
        & 13.6856(6155)
        & 0.9899(2016) \\

    \midrule

    \multirow{3}{*}{Diabetes}
        & $10$ & $4$
        & 0.0003
        & \multicolumn{1}{c}{$<10^{-4}$}
        & 0.0128
        & \multicolumn{1}{c}{$<10^{-4}$}
        & \multicolumn{1}{c}{$<10^{-4}$}
        & 0.0003
        & 0.0294(0091)
        & 8.4878(4520)
        & 3.6482(6539)
        & 0.1348(0103) \\

        & $20$ & $4$
        & 0.0002
        & \multicolumn{1}{c}{$<10^{-4}$}
        & 0.0271
        & \multicolumn{1}{c}{$<10^{-4}$}
        & \multicolumn{1}{c}{$<10^{-4}$}
        & 0.0001
        & 0.0596(0167)
        & 8.2794(4755)
        & 6.6910(7717)
        & 0.1676(0207) \\

        & $20$ & $19$
        & 0.0002
        & \multicolumn{1}{c}{$<10^{-4}$}
        & 0.0041
        & \multicolumn{1}{c}{$<10^{-4}$}
        & \multicolumn{1}{c}{$<10^{-4}$}
        & \multicolumn{1}{c}{$<10^{-4}$}
        & 0.0620(0259)
        & 8.2794(4755)
        & 6.6910(7717)
        & 0.1713(0333) \\

    \bottomrule
  \end{tabular}
\end{table*}

\subsection{Accuracy vs.\ iterations and scale factor}

We compared the output \eqref{eq:privPoe} of Protocol~\ref{alg:poe} with that of the standard distributed GPR in \eqref{eq:DGPR}.
As a performance metric, we consider the root mean square error (RMSE), averaged over all agents:
\begin{equation*}
    \RMSE_{\smash{\hat{f}}}(T) := \frac{1}{M} \sum_{i=1}^M  \sqrt{ \frac{1}{|\calX|} \sum_{x\in\calX}\|\hat{f}(x) - \hat{f}_i^{(T)}(x)\|^2_2}, 
\end{equation*}
where $\calX$ denotes the set of test samples. The metric $\RMSE_{V}(T)$ is defined in a similar manner.
Note that the metric quantifies the discrepancy with respect to \eqref{eq:DGPR} rather than evaluating prediction accuracy, since our objective is to implement \eqref{eq:DGPR} in a privacy-preserving manner, and thus, recover its output.
In this experiment, we varied the number of iterations $T$ and the scale factor $\sfL_z$, while using the same network topology, parameters, and optimized hyperparameters from the previous subsection.
The results in Fig.~\ref{fig:accuracy} show that the RMSE monotonically decreases as $T$ increases and $\sfL_z$ decreases for both datasets, which is consistent with Theorem~\ref{thm:poe}.

\subsection{Total runtime vs.\ network topology}
To investigate the impact of network topology on total runtime, we varied the number of agents $M$ and the number of neighbors per agent, while fixing the number of iterations to $T=100$ and the scale factor to $\sfL_z=1/10^4$.
For practical implementation, instead of executing Protocol~\ref{alg:poe} separately for each test sample (and each output dimension), we stacked all initial values into a single batched vector and performed a single vectorized average consensus. 
This significantly reduces the number of communication rounds and improves efficiency. 
To emulate communication latency between agents, we introduced a fixed delay of $\SI{0.2}{\milli\s}$ per communication, as in the setting of \cite{LuojZhan23}.
All experiments were repeated $20$ times, and Table~\ref{tab:computation_time} reports the mean and standard deviation of the total runtime, excluding hyperparameter optimization and data preprocessing steps.

Table~\ref{tab:computation_time} shows that the total runtime of the proposed protocol increases as either the number of agents or the number of neighbors per agent increases. 
This is primarily due to the increase in the number of messages to be exchanged during the execution of the protocol, as discussed in \eqref{eq:commComplex}.
Nevertheless, it can also be observed that a larger number of neighbors per agent yields a smaller RMSE within the same number of iterations, which may be attributed to the faster convergence of the consensus dynamics on a more densely connected graph.

Next, we compared the proposed protocol with \cite{LuojZhan23,NawaChen24}.\footnote{ Even though \cite{LuojZhan23} and \cite{NawaChen24} were developed for the centralized and the sparse GPR, respectively, both were straightforwardly adapted to \eqref{eq:DGPR}.} 
The former employs additive secret sharing, in which the agents submit shares of their datasets to two computing servers that interact to evaluate the prediction.
The latter employs multi-key HE, in which each agent encrypts its local prediction and interacts with a central server for aggregation and collaborative decryption. 
For the simulation, the multi-key Cheon-Kim-Kim-Song (CKKS) cryptosystem \cite{CheoKima17} was implemented using Lattigo version 6.2.0 \cite{lattigo}, with the same parameters as in \cite{NawaChen24}.
The same kernel and hyperparameters were used, and as neither method is fully distributed, their runtimes depend only on the number of agents and not on the network topology.

Table~\ref{tab:computation_time} shows that the proposed protocol attains a substantially lower total runtime.
This may be attributed to the need to perform matrix inversion over shares required in \cite{LuojZhan23}, which incurs $22N_i-6$ communication rounds for each local dataset, and to the overhead induced by encryption in \cite{NawaChen24}. 
In particular, \cite{LuojZhan23} exceeded $\SI{300}{\second}$ on the SARCOS dataset, and the corresponding entries are omitted.
It is acknowledged that both methods attain a smaller RMSE than the proposed protocol, but this gap can be narrowed by increasing $T$ and decreasing $\sfL_z$.
Overall, these results support the practical applicability of the proposed approach to real-world problems.

\section{Conclusion}\label{sec:conclusion}

In this paper, we proposed a privacy-preserving fully distributed GPR protocol that hides each agent's individual contribution from semi-honest coalitions of bounded size, beyond what is implied by the aggregated value.
By integrating additive secret sharing into a secure distributed average consensus, the protocol guarantees that each agent's local model practically converges to a global model that would be obtained from the standard distributed GPR. 
We analyzed the convergence of the proposed protocol and provided formal privacy guarantees through a simulation based security proof. 
Furthermore, we introduced a method to perform hyperparameter optimization in a privacy-preserving manner by leveraging the proposed protocol. 

\bibliographystyle{IEEEtran}
\bibliography{TNNLS} 

\appendices

\section{Technical Lemmas}\label{apdx:tech}

\begin{lem}\label{lem:tech}
    Consider the symmetric matrix $W$ defined in \eqref{eq:W}. Then, 
\begin{align}\label{eq:Wprop}
    W\bfone = \bfone,~~~~\rho\left(W-\frac{1}{M}\bfone \bfone^\top\right)<1, ~~~~ W\succeq 0.
\end{align}
\end{lem}

\noindent\textit{Sketch of Proof.} 
The first property directly follows from \eqref{eq:wij} and \eqref{eq:W}, and implies that $W$ is doubly stochastic. 
The second property follows from the Perron-Frobenius theorem \cite[Section~8.4]{HornJohn12}, together with the facts that the graph $\calG$ is connected and the diagonal entries of $W$ are positive.
Since $W_{ii}>1/2$ for all $i=1,\ldots, M$, the last property follows from the Gershgorin circle theorem \cite[Section~6.1]{HornJohn12}.
\hfill $\blacksquare$

\section{Proof of Theorem~\ref{thm:metroQuant2}}\label{apdx:metroQuant2}

In what follows, we show that 
\begin{align*}
    \Big\|\sum_{j\in\calN_i} \bar{w}_{ij}\left(\sfQ(z_j(t)) - \sfQ(z_i(t)) \right) \Big\|<\frac{q}{2},
\end{align*}
for all $t\ge 0$ and $i\in\calV$ under \eqref{eq:qbound}. 
Given this, since $a=a\modp q$ for all $a\in\bbZ$ such that $|a|<q/2$, the dynamics \eqref{eq:metroQuant} can be rewritten as 
\begin{align}\label{eq:metroQuantRe}
    &\!\!z_i(t+1) = z_i(t) + \sfL_z\sfL_w \!\sum_{j\in\calN_i}\!  \bar{w}_{ij}\left(\sfQ(z_j(t)) - \sfQ(z_i(t)) \right) \\
    &\!\!= z_i(t) +\sfL_z\sfL_w \left(\sum_{j\in\calN_i} \bar{w}_{ij}\left(\sfQ(z_j(t)) - \sfQ(z_i(t)) \right) \modp q \right)\!, \nonumber  
\end{align}
which is equivalent to the form of \eqref{eq:metroFinal} subject to \eqref{eq:zeroshare}.
Hence, $z_i(t)\equiv \bar{z}_i(t)$ for all $i\in\calV$, and \eqref{eq:thmFinalMetroToShow} follows from Lemma~\ref{lem:quantMetro}.

Using \eqref{eq:W}, the dynamics \eqref{eq:metroQuant} can be compactly rewritten as
% We begin by compactly rewriting the dynamics \eqref{eq:metroQuant} as
% \begin{align}\label{eq:KronRe}
%     &z(t+1) \\
%     &= \left(W \otimes I_p \right) z(t) + \left(\left(W-I \right) \otimes I_p \right) \left( \sfL_z \cdot \sfQ(z(t)) - z(t)\right), \nonumber
% \end{align}
\begin{align}\label{eq:KronRe}
    z(t+1) = \left(W \otimes I_p \right) z(t) + \left(\left(W-I \right) \otimes I_p \right) g(z(t)),
\end{align}
where $z(t):=[z_1(t);\cdots;z_M(t)]\in\bbR^{Mp}$, $\sfQ(z(t)):=[\sfQ(z_1(t));\cdots;\sfQ(z_M(t))]\in\bbR^{Mp}$, 
$g(z(t)):= \sfL_z\cdot \sfQ(z(t)) - z(t)$,
and $\otimes$ denotes the Kronecker product.

To separate the \textit{average} component from the \textit{disagreement} component, we introduce a matrix $R\in \bbR^{M \times (M-1)}$ whose columns form an orthonormal basis of $\ker(\bfone^\top)$, so that
\begin{align}\label{eq:Rprop}
    R^\top R = I_{M-1}, ~~R^\top \bfone  = \bfzero, ~~RR^\top = I_M - \frac{1}{M}\bfone \bfone^\top,
\end{align}
and define the coordinate transformation 
\begin{subequations}
\begin{align}
    \begin{bmatrix}
        \calZ_o(t) \\
        \tilde{\calZ}(t)
    \end{bmatrix}
    &:= \begin{bmatrix}
        \frac{1}{M} \bfone^\top \otimes I_p \\
        R^\top \otimes I_p 
    \end{bmatrix} z(t), \\
    z(t) &= (\bfone \otimes I_p) \calZ_o(t) + \left( R \otimes I_p \right) \tilde{\calZ}(t). \label{eq:transformInv}
\end{align}
\end{subequations}
Here, $\calZ_o(t) \in \bbR^p$ is the average of the states $z_i(t)$, and $\tilde{\calZ}(t)\in\bbR^{(M-1)p}$ quantifies their disagreement in the sense that $\tilde{\calZ}(t) = \bfzero$ if and only if all $z_i(t)$ are identical.

Applying the transformation to \eqref{eq:KronRe} and using \eqref{eq:Wprop} together with $(A\otimes B)(C\otimes D) = (AC)\otimes(BD)$, we obtain
\begin{align*}
    \calZ_o(t+1) &= \left(\frac{1}{M} \bfone^\top \otimes I_p \right) \left(W\otimes I_p \right) z(t) \\
    &~~ + \left(\frac{1}{M} \bfone^\top \otimes I_p \right) \left(\left(W-I \right) \otimes I_p \right) g(z(t)) \\
    &=\left(\frac{1}{M} \bfone^\top \otimes I_p \right) z(t) = \calZ_o(t).
\end{align*}
Hence, the average component is invariant, i.e.,
$\calZ_o(t) \equiv z^\avg$.

For the disagreement component, first observe that
\begin{align}\label{eq:Lambda}
    \begin{bmatrix}
        \frac{1}{M} \bfone^\top  \\
        R^\top 
    \end{bmatrix} W 
    \begin{bmatrix}
        \bfone   & R  
    \end{bmatrix} = 
    \begin{bmatrix}
        1 & 0 \\
        0 & \Lambda
    \end{bmatrix},
\end{align}
by \eqref{eq:Wprop} and \eqref{eq:Rprop},
where $\Lambda := R^\top W R\in \bbR^{(M-1)\times (M-1)}$.
Since $R^\top W = R^\top W (RR^\top + \frac{1}{M}\bfone \bfone^\top ) = \Lambda R^\top$,
we have
\begin{align}\label{eq:tildez}
    &\tilde{\calZ}(t+1) = \left( R^\top \otimes I_p \right)\left( W \otimes I_p \right) z(t)  \\
    & ~~+ \left( R^\top \otimes I_p \right) \left( (W-I)\otimes I_p \right) g(z(t)) \nonumber\\
    &= \left( \Lambda \otimes I_p \right)\left( R^\top\otimes I_p \right) z(t)  + \left( \left(\Lambda - I \right)R^\top \otimes I_p \right) g(z(t)) \nonumber\\
    &= \left( \Lambda \otimes I_p \right) \tilde{\calZ}(t) \! + \!\left( \left(\Lambda \!-\! I \right)  R^\top \!\otimes I_p \right) \! g(z(t)) . \nonumber
\end{align}

The block diagonalization in \eqref{eq:Lambda} implies that the eigenvalues of $\Lambda$ are those of $W$ except the eigenvalue at $1$ associated with the eigenvector $\bfone$.  
Since $\frac{1}{M}\bfone \bfone^\top$ is the orthogonal projection onto $\bfone$, subtracting it from $W$ replaces this eigenvalue at 1 with 0, and leaving the others unchanged. Hence, because $\Lambda$ is symmetric,
\begin{align*}
    \|\Lambda\|_2 = \rho(\Lambda) =  \rho\left(W - \frac{1}{M}\bfone \bfone^\top\right) =\lambda<1, 
\end{align*}
where $\|\cdot\|_2$ is the (induced) 2-norm.
Moreover, by Lemma~\ref{lem:tech}, the eigenvalues of the symmetric matrix $\Lambda - I$ lie in $[-1,\lambda-1]$, and thus, $\|\Lambda-I\|_2 =\rho(\Lambda-I) \le 1$.
Using $\|A\otimes B\|_2 = \|A\|_2\cdot \|B\|_2$ and $\|R^\top\|_2 = 1$ leads to
\begin{align*}
    \left\| \tilde{\calZ}(t+1) \right\|_2 & \le \left\| \Lambda \right\|_2  \left\| \tilde{\calZ}(t) \right\|_2 + \left\| \Lambda - I\right\|_2  \left\| g(z(t)) \right\|_2 \\
    &\le \lambda \left\| \tilde{\calZ}(t) \right\|_2 + \left\| g(z(t)) \right\|_2.
\end{align*}
By the definition of the quantizer in \eqref{eq:Lz}, it holds that $\| g(z(t)) \|_2 \le \sfL_z\sqrt{Mp}/2$ for all $z(t)\in\bbR^{Mp}$.
Therefore, we can further obtain that 
\begin{align*}
    \left\| \tilde{\calZ}(t) \right\|_2 & \le \lambda^t \left\| \tilde{\calZ}(0) \right\|_2 + \frac{\sfL_z\sqrt{Mp}}{2} \sum_{k=0}^{t-1} \lambda^k \\
    & \le \left\| \tilde{\calZ}(0) \right\|_2 + \frac{\sfL_z\sqrt{Mp}}{2(1-\lambda)}  \le \sqrt{Mp} \tilde{z}_{\max}^{\ini} + \frac{\sfL_z\sqrt{Mp}}{2(1-\lambda)},
\end{align*}
where the last inequality follows from \eqref{eq:transformInv} and $\calZ_o(0)=z^\avg$:
\begin{align*}
    \left\|\tilde{\calZ}(0)\right\|_2 & = \left\| z(0) - \left(\bfone \otimes I_p \right) \calZ_o(0) \right\|_2 \\
    &= \left( \sum_{i=1}^M \left\| z_i^\ini - z^\avg \right\|_2^2 \right)^{1/2} \le \sqrt{Mp} \tilde{z}_{\max}^\ini.
\end{align*}

Finally, since $\|\cdot\|\le \|\cdot\|_2$ for vectors and $\|R\otimes I_p\|_2 =1$, \eqref{eq:transformInv} with $\calZ_o(t)\equiv z^\avg$ gives $\| z(t) \| \le \| z^\avg \| + \|\tilde{\calZ}(t) \|_2$.
Moreover, it is clear from \eqref{eq:wij} and \eqref{eq:Lw} that $\max_{i,j\in\calV} \|\bar{w}_{ij}\| \le 1/4\sfL_w$.
Consequently,
\begin{align*}
    &\Big\| \sum_{j\in\calN_i} \bar{w}_{ij} \left( \sfQ(z_j(t)) - \sfQ(z_i(t)) \right) \Big\| \\
    &\le M\cdot \max_{i,j\in\calV} \left\| \bar{w}_{ij} \right\| \cdot \max_{i,j\in\calV} \left\| \sfQ(z_j(t)) - \sfQ(z_i(t)) \right\| \\
    & \le \frac{M}{4\sfL_w} \cdot 2\left\| \sfQ(z(t)) \right\|  \le  \frac{M}{2\sfL_w} \left( \frac{1}{2} + \frac{\|z(t)\|}{\sfL_z} \right) \\
    & \le \frac{M}{2\sfL_w} \left( \frac{1}{2} + \frac{\|z^\avg\| + \sqrt{Mp}\tilde{z}_{\max}^\ini }{\sfL_z} + \frac{\sqrt{Mp}}{2(1-\lambda)}  \right) < \frac{q}{2},
\end{align*}
and this concludes the proof.
\hfill $\blacksquare$

\section{Proof of Theorem~\ref{thm:priv}}\label{apdx:priv}

We focus on the case $T=1$ because Protocol~\ref{alg:metro} with an arbitrary $T\in\bbN$ can be realized as a sequential composition of Protocol~\ref{alg:metro} with $T=1$, and such composition preserves security \cite{Cane00}. 
Accordingly, we set $\sfy_i=\bar{z}_i(1)$. 
In the remainder of the proof, we suppress the time index, e.g., writing $s_{j\to i}^i$ for $s_{j\to i}^i(0)$, for brevity.

The collection of all messages received by agent $i$ during the execution of the protocol is given by 
\begin{align*}
    \msg_i \!=\! \Big(
\left\{s_{j\to i}^i \right\}_{j\in\calN_i},
\left\{\zeta_{ij} \right\}_{j\in\calN_i},
\left\{s_{l\to i}^j \right\}_{j\in\calN_i, l\in\calN_i\cap\calN_j^+}
\Big)\!.
\end{align*}
The first two components correspond to the shares and masked messages that agent $i$ receives as an aggregator, respectively.
The last component represents the shares it receives when its neighbor $j$ is the aggregator.

Since \eqref{eq:jshare} and \eqref{eq:ishare} split the zero vector into additive shares, any single share can be uniquely determined by the remaining shares.
To prevent the coalition from exploiting this, we require $|\calC|\le \min_{(i,j)\in\calE}|\calN_i^+ \cap \calN_j^+|-2$. 
Then, for every $i\in\calV$ and $j\in\calN_i$, there exist at least two non-colluding agents $l_1,l_2 \in (\calN_i^+\cap\calN_j^+)\setminus \calC$. 
Consequently, the shares exchanged between $l_1$ and $l_2$ cannot be reconstructed by the coalition due to the randomness of the share generation algorithm. 
In other words, the coalition cannot determine any share that it does not receive, and thus, all components of $(\msg_i)_{i\in\calC}$ are uniformly distributed in $\bbZ_q^p$ from the coalition's view, subject only to the constraints \eqref{eq:metroFinal} and \eqref{eq:mask}. 

In this regard, for each coalition $\calC\subset \calV$, we construct a simulator $\Sim_\calC$ that outputs 
\begin{multline*}
\Sim_\calC((\sfx_i,\sfy_i)_{i\in\calC}) 
= 
\Big(
\sfx_i, 
\left\{\hat{s}_{j\to i}^i \right\}_{j\in\calN_i^+},   
\left\{\hat{\zeta}_{ij} \right\}_{j\in\calN_i}, \\
\left\{\hat{s}_{l\to i}^j\right\}_{j\in\calN_i, l\in\calN_i^+\cap\calN_j^+} 
\Big)_{i\in\calC},
\end{multline*}
and emulates as follows for each $i\in\calC$:
\begin{enumerate}
    \item For all $j\in\calN_i$, sample $\hat{\zeta}_{ij}$ from $\bbZ_q^p$ uniformly at random.
    \item For all $j\in\calN_i$ and $l\in\calN_i^+ \cap \calN_j^+$, sample $\hat{s}^j_{l\to i}$ from $\bbZ_q^p$ uniformly at random.
    \item Using $(\sfx_i,\sfy_i)$ and the chosen $\{\hat{\zeta}_{ij}\}_{j\in\calN_i}$, compute $\hat{\phi}_{ii}\in\bbZ_q^p$ that satisfies
    \begin{align*}
        \sfy_i \!&=\! z_i^\ini \!+\! \sfL_w\sfL_z  \Big( \hat{\phi}_{ii} + \!\! \sum_{j\in\calN_i} \hat{\zeta}_{ij} - \bar{w}_{ij}\sfQ(z_i^\ini) \modp q \Big)
    \end{align*}
    \item Generate
    \begin{align*}
        (\hat{s}_{j\to i}^i)_{j\in\calN_i^+}\leftarrow \Share(\hat{\phi}_{ii}, |\calN_i^+|),
    \end{align*}
    so that $\hat{\phi}_{ii} = \sum_{j\in\calN_i^+} \hat{s}_{j\to i}^i \modp q$ holds.

\end{enumerate}

By construction, each $\hat{s}_{j\to i}^i$, $\hat{\zeta}_{ij}$, and $\hat{s}_{l\to i}^j$ is uniformly distributed in $\bbZ_q^p$ and satisfies the same constraints \eqref{eq:metroFinal} and \eqref{eq:mask}.
Therefore, \eqref{eq:privGoal} holds, and this concludes the proof.
\hfill $\blacksquare$

\section{Proof of Theorem~\ref{thm:poe}}\label{apdx:poe}
    Let us fix an index $i\in\calV$.
    By Theorem~\ref{thm:metroQuant2}, the output $\bar{z}_i(T)$ of Protocol~\ref{alg:metro} with \eqref{eq:init} satisfies
    \begin{equation}\label{eq:thmPoeProp}
    \!\!\limsup_{T\to\infty } \left\| 
    \begin{bmatrix}
        \bar{z}_{i,1}(T) \\ \bar{z}_{i,2}(T)       
    \end{bmatrix} \!-\!
    \begin{bmatrix}
        \sum_{l=1}^M V_l^{-1}(x) \cdot \hat{f}_l(x) \\
        \sum_{l=1}^M V_l^{-1}(x)
    \end{bmatrix}  \right \| \le \sfL_z.
    \end{equation} 
    Then, by \eqref{eq:xi}, for any $\epsilon_0\in(0,1/V(x)-\sfL_z)$ there exists $T_0\ge0$ such that $\| 1/V(x) - \bar{z}_{i,2}(T)\| \le \sfL_z + \epsilon_0 $ for all $T\ge T_0$. 
    For such $T$, $\bar{z}_{i,2}(T) \ge 1/V(x)-\sfL_z-\epsilon_0>0$, and thus, we have
    \begin{align*}
        &\left\| V^{(T)}_i(x) - V(x) \right\| 
        = \left\| \frac{1/V(x) - \bar{z}_{i,2}(T)}{\bar{z}_{i,2}(T)/V(x)}  \right\| \\
        &\le \frac{\sfL_z+\epsilon_0}{(1/V(x) - \sfL_z-\epsilon_0)/V(x)} = \gamma(\sfL_z+\epsilon_0).
    \end{align*} 
    Letting $\epsilon_0 \to 0$ and using continuity of $\gamma$, which is a class $\calK$ function under \eqref{eq:xi}, satisfies \eqref{eq:thmPoeToShow2}.
    
    Next, by \eqref{eq:thmPoeProp}, for the same $\epsilon_0$ there exists $T_1 \ge 0$ such that
    $\| \bar{z}_{i,1}(T) -\sum_{l=1}^M V_l^{-1}(x)\cdot\hat{f}_l(x) \| \le \sfL_z+\epsilon_0$ for all $T \ge T_1$,
    which implies $\| \bar{z}_{i,1}(T) \| \le \| \sum_{l=1}^M V_l^{-1}(x)\cdot\hat{f}_l(x) \| + \sfL_z + \epsilon_0$.
    Then, for all $T\ge \max\{T_0,T_1\}$, it follows that
    \begin{align*}
    &\left\| \hat{f}^{(T)}_i(x) -\hat{f}(x) \right\| \le \left\| \left(V^{(T)}_i(x) - V(x) \right) \bar{z}_{i,1}(T)\right\| \\
    &~~~~ + \left\| V(x)\left( \bar{z}_{i,1}(T) -\sum_{l=1}^M V_l^{-1}(x)\cdot\hat{f}_l(x)  \right) \right\| \\
    &\le \gamma(\sfL_z+\epsilon_0) \cdot \left(\left\| \sum_{l=1}^M V_l^{-1}(x)\cdot\hat{f}_l(x) \right\| + \sfL_z +\epsilon_0 \right) \\
    & ~~~~ + (\sfL_z+\epsilon_0)\cdot V(x) =: \beta(\sfL_z+\epsilon_0),
    \end{align*}
    where we used the triangle inequality and \eqref{eq:thmPoeToShow2}. 
    The function $\beta$ is clearly of class $\calK$ and continuous.
    Hence, letting $\epsilon_0\to 0$  establishes \eqref{eq:thmPoeToShow1}, and this concludes the proof. \hfill $\blacksquare$

\vskip -2.5\baselineskip plus -1fil

\begin{IEEEbiography}[{\includegraphics[width=1in,height=1.25in,clip,keepaspectratio]{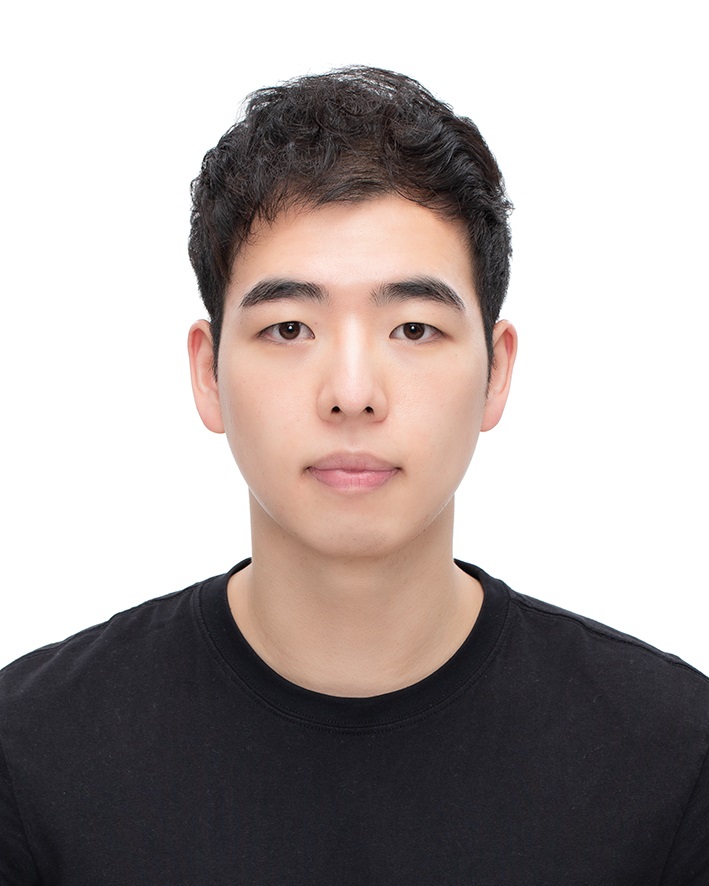}}]%
{Yeongjun Jang}
received the B.S. degree in electrical and computer engineering from Seoul National University, South Korea, in 2022.
He is currently a combined M.S./Ph.D. student in electrical and computer engineering at Seoul National University, South Korea. 
His research interests include control system security and stochastic optimal control.
\end{IEEEbiography}
\vskip -2.5\baselineskip plus -1fil

\begin{IEEEbiography}[{\includegraphics[width=1in,height=1.25in,clip,keepaspectratio]{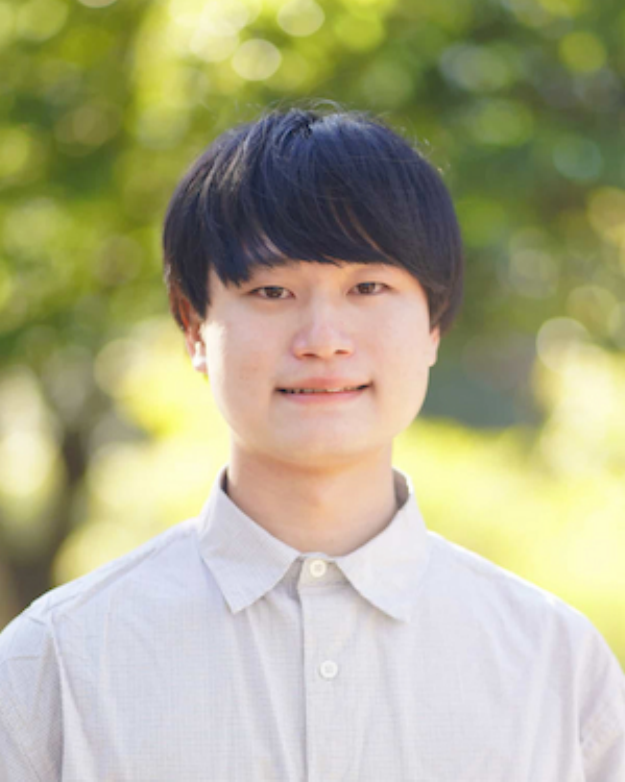}}]%
{Kaoru Teranishi}
received the B.S. degree in electronic and mechanical engineering from the National Institute of Technology, Ishikawa College, Ishikawa, Japan, in 2019, and the M.S. and Ph.D. degrees in mechanical and intelligent systems engineering from the University of Electro-Communications, Tokyo, Japan, in 2021 and 2024, respectively.
From 2019 to 2020, he was a Visiting Scholar at the Georgia Institute of Technology, GA, USA.
From 2021 to 2024, he was a Research Fellow at the Japan Society for the Promotion of Science (JSPS), Tokyo, Japan.
From 2024 to 2025, he was a JSPS Overseas Research Fellow.
During this period, he was a Postdoctoral Research Affiliate at The University of Texas at Austin, TX, USA, and a Visiting Scholar/Post-Doctoral Research Associate at Purdue University, IN, USA.
Since December 2025, he has been an Assistant Professor in the Department of Information and Physical Sciences, Graduate School of Information Science and Technology, The University of Osaka.
His research interests include control theory and cryptography for control systems security.
\end{IEEEbiography}
\vskip -2.5\baselineskip plus -1fil

\begin{IEEEbiography}[{\includegraphics[width=1in,height=1.25in,clip,keepaspectratio]{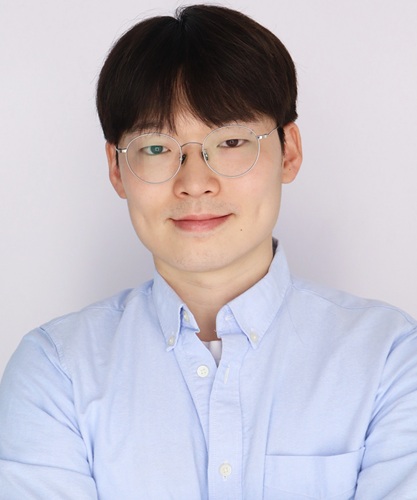}}]%
{Jihoon Suh}
received the B.S. degree in aerospace engineering in 2018 and the M.S. degree in Control Theory in 2020, from The University of Texas at Austin, TX, USA. 
He received the Ph.D. degree from the School of Aeronautics and Astronautics at Purdue University, West Lafayette, IN, USA, in 2026. His research interests include control theory, reinforcement learning, encrypted control and optimization. He is currently a research scientist at Anduril Industries.
\end{IEEEbiography}
\vskip -2.5\baselineskip plus -1fil

\begin{IEEEbiography}[{\includegraphics[width=1in,height=1.25in,clip,keepaspectratio]{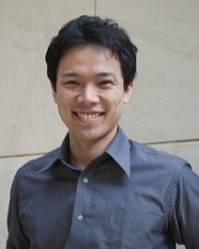}}]%
{Takashi Tanaka}
received the B.S. degree from the University of Tokyo, Tokyo, Japan, in 2006, and the M.S.
and Ph.D. degrees in aerospace engineering (automatic control) from the University of Illinois at Urbana Champaign, Champaign, IL, USA, in 2009 and 2012, respectively. 
He was a Postdoctoral Associate with the Laboratory for Information and Decision Systems at the Massachusetts Institute of Technology, Cambridge, MA, USA, from 2012 to 2015, and a postdoctoral researcher at KTH Royal Institute of Technology, Stockholm, Sweden, from 2015 to 2017. 
He was an Assistant Professor in the Department of Aerospace Engineering and Engineering Mechanics at the University of Texas at Austin between 2017 and 2024, and is an Associate Professor at the School of Aeronautics and Astronautics and the Elmore Family School of Electrical and Computer Engineering at Purdue University since 2025.
His research interests include control theory and its applications, most recently the information-theoretic perspectives of optimal control problems. 
He was the recipient of the DARPA
Young Faculty Award, the AFOSR Young Investigator Program Award, and the NSF Career Award.
\end{IEEEbiography}

\end{document}